\documentclass{article}
\usepackage[utf8]{inputenc}
\usepackage{graphicx} 
\usepackage{subcaption} 
\usepackage{float} 
\usepackage{tikz} 
\usepackage{booktabs}
\usepackage{svg}
\usepackage{makecell}
\usepackage{lmodern}  
\usepackage{babel}    
\usepackage{comment} 
\usepackage{amsmath}
\usepackage{amsfonts}
\usepackage{cancel} 
\usepackage{amssymb}
\usepackage{lmodern}
\DeclareSymbolFont{matha}{OML}{txmi}{m}{it}
\DeclareMathSymbol{\varv}{\mathord}{matha}{118}
\usepackage{mathtools} 
\usepackage[version=4]{mhchem} 
\usepackage{multicol} 
\usepackage{hyperref} 
\usepackage{xcolor} 

\usepackage{hyperref}
\hypersetup{
    colorlinks = true,      
    linkcolor  = blue,      
    citecolor  = blue,      
    urlcolor   = cyan       
}
\usepackage{xurl} 
\usepackage{orcidlink} 
\usepackage{cleveref} 
\usepackage[numbers]{natbib} 
\usepackage{geometry}
 \usepackage{authblk} 

 \usepackage[utf8]{inputenc}

 \usepackage{float}

\usepackage{titlesec}
\titleformat{\section}
  {\large\bfseries} 
  {\thesection}
  {1em}
  {}

\usepackage{caption}

\usepackage[normalem]{ulem}
\definecolor{ADdel}{RGB}{190,25,25}
\definecolor{ADadd}{RGB}{0,70,190}
\definecolor{ADnote}{RGB}{200,95,0}

\title{\textbf{Neutron-Lean D-3He Fusion by Asymmetric Enhancement of the Primary Reaction: A Spin-Polarized Non-Thermal Hot Tail}}

\author[1]{Pascal Zhang\,\orcidlink{0009-0007-3709-1785}}
\author[1,2]{Ahmed Diallo\,\orcidlink{0000-0002-0706-060X}}      

\affil[1]{Princeton Plasma Physics Laboratory, Princeton University, Princeton, NJ, USA}
\affil[2]{Now at Proxima Fusion, Munchen, Germany}

\date{}

\DeclareUnicodeCharacter{2009}{\,}
\begin{document}

\maketitle

E-mail : \href{mailto:zhangpascal0@gmail.com}{\texttt{zhangpascal0@gmail.com}}, \href{mailto:adiallo@proximafusion.com}{\texttt{adiallo@proximafusion.com}}

\vspace{1em}
\hrule 
\vspace{1em}

\noindent \textbf{Abstract:} 
Advanced low-neutron-yield fusion fuels, such as $\ce{D-^3He}$, offer a solution to the severe material degradation and neutron activation inherent in standard $\ce{D-T}$ reactors. However, the $\ce{D-^3He}$ reaction in a thermal regime has a very high ideal ignition temperature (bremsstrahlung crossover temperature) on the order of $30~\text{keV}$ and an ideal operating temperature on the order of $70~\text{keV}$, whereas conventional $\ce{D-T}$ reactors have an ideal ignition temperature on the order of $4~\text{keV}$ and an operating temperature between $10~\text{keV}$ and $20~\text{keV}$. Furthermore, $\ce{D-^3He}$ fuel cannot be considered completely aneutronic since it presents secondary $\text{D--D}$ reactions that produce neutrons. This paper focuses on a modified fuel that combines the theory of spin-polarized fuel with non-thermal ``hot-tail'' reactions for $\ce{D-T}$, $\ce{D-^3He}$, and $\text{D--D}$ reactions. We find that this combination is particularly advantageous for the $\ce{D-^3He}$ fuel in several respects. First, a large reactivity enhancement :one order of magnitude at $10~\text{keV}$ for a small hot tail ($F=0.1$), rising to two orders of magnitude for a larger one ($F=0.6$), with polarization adding a further factor of $1.5$. Second, an ideal ignition energy far below the thermal value: $9.5~\text{keV}$ at $F=0.3$ down to $3.2~\text{keV}$ at $F=0.5$, which polarization reduces further to $3.8~\text{keV}$ and $1.3~\text{keV}$, respectively. Third, the asymmetric enhancement between the primary reaction and the secondary reactions in the fuel opens the possibility of a neutron-lean reaction. Quantifying this with a neutrons-per-watt metric, we find that a thermal $\ce{D-^3He}$ plasma at $10~\text{keV}$ is barely neutron-lean at all ($90$--$95\%$ of the D--T neutron rate per watt), whereas a hot tail of $F=0.1$ reduces it by $4.7$--$7.7\times$ and the polarized non-thermal fuel at its ignition point reaches $3.0$--$5.4\%$ of the D--T value, with only $0.4$--$2.4\%$ of the fusion power carried by neutrons. The range spans prompt removal versus complete in-situ burnup of the bred tritium. Applied to a $200~\text{MW}$ worked example anchored on the ARTEMIS $\ce{D-^3He}$ reactor design, this is a first-wall neutron loading of $0.008$--$0.043~\text{MW\,m}^{-2}$ against $1.45~\text{MW\,m}^{-2}$ for D--T in the same chamber, extending the time to reach $20~\text{dpa}$ from about $1.5$ full-power years to $27$.
\vspace{1em}
\hrule 
\vspace{1em}

\begin{flushleft}
\end{flushleft}

\tableofcontents

\begin{multicols}{2}
\section{Introduction}
\end{multicols}

\begin{align}
& D \ + \ \  D   \ \ \ \to T \ (1.01 \ MeV) \ + \ p \ (3.03 \ MeV) \quad \  Q_{DD_p} \ \ \  = 4.04 \ MeV \label{ddp} \\
& D \ + \ \  D   \ \ \ \to \ce{^3He} \ (0.82 \ MeV) \ + \ n \ (2.45 \ MeV) \ \  Q_{DD_n} \ \ \  = 3.27 \ MeV \label{ddn} \\
& D \ + \ \  T   \ \ \ \to \alpha \ (3.52 \ MeV) \ + \ n \ (14.06 \ MeV) \quad Q_{DT} \ \ \  = 17.58 \ MeV \label{DT} \\
& D  \ +  \ ^{3}He\to \alpha  \ (3.67 \ MeV)  \ + \ p \ (14.67 \ MeV) \quad Q_{\ce{D^{3}He}}=18.34 \ MeV  \label{DHe3}
\end{align}

\begin{multicols}{2}

\par The fusion reaction between deuterium and tritium (D--T) is widely favoured due to its high fusion probability and an ideal ignition temperature of 4.3 keV \cite{Book_Bremsshalung} (Ideal reactor where losses are only from bremsstrahlung radiation), which is significantly lower than that of other fusion reactions. However, this fusion reaction produces a high-energy neutron of 14 MeV \cref{DT} that can cause severe damage by displacing atoms in the crystal structure \cite{Displacement} and activating the materials of the fusion device. This activation has a twofold effect: it produces short-lived radioactive waste (with half-lives of 80 to 100 years) \cite{Materials} and makes maintenance work near to the reactor impossible, as it could expose personnel to radiation.

\par That is why we have begun to explore  fuels that do not produce neutrons, known as aneutronic fuels which are also referred to as advanced fuels or second-generation fuels \cite{Advanced_fuel,Advanced_fuel_2,Advanced_fuel_3}. One of the first candidates  was the $\ce{D-^3He}$ reaction \cref{DHe3}, even though the fuel itself cannot be considered completely aneutronic due to D--D side reactions \cref{ddn,ddp}. However, this fuel produce significantly fewer neutrons than the D--T fuel. Nevertheless, what truly hinders the development of a $\ce{D-^3He}$ fuel lies in the physics itself: the enormous difficulty in igniting this reaction due to its low fusion probability at temperatures comparable to those of the D--T reaction, which results in an ideal minimum startup temperature of around 30 keV \cite{Book_Bremsshalung}, temperature that is not suitable for producing usable power, which requires a temperature in reality of around 70 keV \cite{Feasibility_Advanced_fuel}.  The other major challenge with this fuel is the supply of helium-3, which is present only in very small quantities on Earth, as it is a stable isotope of helium produced from the decay of tritium. it is present in reasonable quantities on the Moon, which raises the possibility of a hypothetical exploitation of lunar soil \cite{Advanced_fuel_3}. 

\par Another advantage of the $\ce{D-^3He}$ reaction would be the direct conversion of the charged particles produced by the reaction into electrical power; it is estimated that this type of conversion would have an efficiency of about 80 \% \cite{Artemis_DC}, whereas the efficiency of the D--T reaction is only about one-third, achieved through a turbo-generator system that converts the thermal power of the neutrons produced, into electrical power.

\par The non-thermal approach rests on the observation that any distribution displaced from a Maxwellian to higher energy, while conserving the total number of particles and the total energy, will enhance the reactivity of the reaction. Such departures already occur in practice: instabilities, auxiliary heating and other transient processes in a fusion device continually push the ion velocity distribution slightly away from an ideal Maxwellian distribution. Non thermal fusion distribution like drift-ring beam, slowing-down, kappa super-thermal and bi-Maxwellian have been studied years ago \cite{bi_maxwellian,Kong_non_thermal,upper_bound_non_thermal}  and some non-thermal fusion reaction have been observed into Jet device \cite{Jet_non_thermal} and into TAE FRC device \cite{TAE_non_thermal}. Researchers at TAE Technologies reported the first direct observation of a beam-driven wave directly energizing plasma fuel ions in a field-reversed configuration (FRC) fusion device (the C-2U experiment). Driven by high-power neutral beam injection (NBI) in the open field line mirror region, this wave creates a high-energy "tail" of fuel ions on sub-collisional timescales. This non-thermal ion acceleration bypasses standard electron friction, dramatically boosting the deuterium-deuterium (D–D) fusion reaction rate without degrading global plasma confinement. Based on this experimental observation Nicolas (2026) \cite{Nicolas}, developed a non-thermal model that is in particular highly advantageous in terms of reactivity enhancement for the $\ce{D-^3He}$ reaction in the range of fusion relevant temperatures (10–20 keV). However these enhancements were argued to be inaccessible, because the $\ce{D-^3He}$ thermal ideal ignition temperature is around 30 keV, except if electrons could be slowing down, decreasing the ignition temperature. 

\begin{figure*}[htbp]
    \centering
    \phantomsection
    \includegraphics[width=1\linewidth]{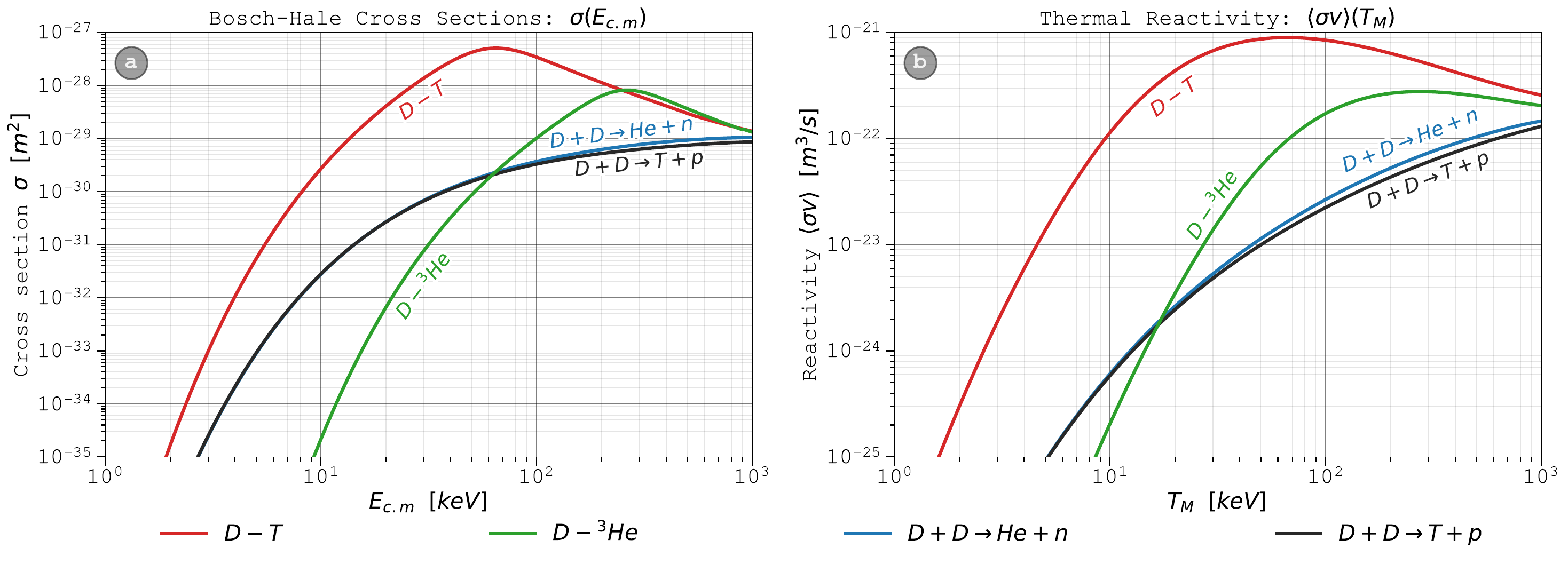}
    \caption{Cross section (a) and thermal reactivity (b)  for D-T, $\ce{D-^3He}$ and D-D reactions computed from Bosch and Hale formula \cite{Bosch_and_Hale}}
    \label{Thermal}
\end{figure*}

\par It is well established that controlling the spin direction of the nuclei involved in D--T and $\ce{D-^3He}$ fusion reaction can lead to an increase of the nuclear cross section by a factor of 1.5. It has been theoretically proven by Kulsrud et al \cite{Kulsrud,Kulsrud_Polarized_fusion_reactor,Polfusion_polarize_fuel} and thus ultimately increase the reactivity of these two reactions by a factor of 1.5. This is therefore our second step  toward improving the reactivity of these reactions, but it is not without its challenges. Indeed, given the complex environment (i.e Tokamak) in which these polarized species exist, many interactions can lead to depolarization \cite{Kulsrud,Depolarization, Lifetime_DT,Persistence_DT,Persistence_of_the_polarization_in_fusion_process} and then cancel its effects on the one hand, and on the other hand, there are challenges related to the methods for producing this polarized fuel\cite{Kulsrud}.
\par However, polarization does not necessarily result in an increase in the cross section for all fusion reactions, and this may be the case for the D--D reactions \cite{Book_Polarized_Fuel}. The various theories regarding the polarization of D--D reactions do not point to a clear trend, some predict a decrease in the cross section for both D--D branches \cite{Doolen,Deltuva,Viviani}, while others predict, on the contrary \cite{Lemaitre}, similarly to the D--T and $\ce{D-^3He}$ reactions there would also be an increase in the cross section. The lack of data currently makes it impossible to determine which theory best reflects reality.  Currently, several experiments—notably the PolFusion experiment, which aims to gather data in the 10–100 keV range is underway \cite{The_status_of_double_polarized_dd,View_on_Polfusion,Optimisation_Pol_Fusion}. Depending on the results obtained, If a clear decrease in the cross section of the D--D reaction is demonstrated, polarization could favour the \ce{D-^3He} and then the possibility of a neutron lean reactor could be explored.

Our paper is organized as follows: In Section \labelcref{sec:Thermal fusion reaction}, we define the nuclear cross section and then the reactivity of four fusion reactions : D--T, $\ce{D-^3He}$, and the two D—D branches under the Maxwellian thermal regime. In Section \labelcref{sec:Non-thermal fusion reaction}, we will define our non-thermal regime and examine its impact on the reactivity of these four reactions, as well as on a $\ce{D-^3He}$ fuel specifically. Subsequently, in Section \labelcref{sec:Polarized non-thermal fusion reaction}, we introduce a polarization into our reactions with the non-thermal regime and as well examines its impact on the reactivity and on a $\ce{D-^3He}$ fuel specifically. Finally, in Section \labelcref{sec:Ignition energy}, we attempt to estimate the ideal ignition energy for the non-thermal fusion regime then for the polarized non-thermal regime of the D--T and $\ce{D-^3He}$ reactions.

\section{Thermal fusion reaction} \label{sec:Thermal fusion reaction}

\par The probability that a nuclear reaction will occur at a specific energy $E$ is quantified  by the nuclear cross section $\sigma(E)$ which has the dimensions of an area. The cross section depends on the reactants
(deuterium, tritium, helium-3), which set the Coulomb barrier and hence the quantum-tunnelling probability. The nuclear cross section for fusion reaction is displayed in \cref{Thermal}.a for four different reactions : D-D (\cref{ddp} and \cref{ddn}), D--T (\cref{DT}) and $\ce{D-^3He}$ (\cref{DHe3}), using Bosch and Hale parametrisation \cite[eq.8, eq.9, table IV, table VI]{Bosch_and_Hale}. These computed nuclear cross sections in \cref{Thermal}.a are all valid in the shared range of energy between 0.5 keV 4700 keV (given in $E_{c.m}$), these bounds are respected throughout. The centre-of-mass energy $E_{c.m}$ is given by $E_{c.m}=\frac{1}{2} \mu u^2$ with $\mu=\frac{m_1m_2}{m_1+m_2}$ the reduced mass and $u=| \vec{v_2}-\vec{v_1}|$ the relative speed.

\par The D--T reaction has the largest cross section over a wide range of low energies, and its cross section peaks at 64 keV (see \cref{table Ed}) because of the resonant intermediate helium-5 state  \cite{Kulsrud}. This peak value is larger than the peak of the $\ce{D-^3He}$ cross section, which occurs near 250 keV.

\par In a working reactor the ions are not mono-energetic: at temperature $T_M$ (temperatures are expressed in energy units throughout, $k_B=1$) their velocities follow a Maxwell--Boltzmann distribution of thermal speed $v_{th}=\sqrt{2T_M/m}$. The macroscopic fusion rate is set by the reactivity $\langle\sigma v\rangle$ $[\mathrm{m^{3}\,s^{-1}}]$, the cross section averaged over the relative-velocity distribution of the two reactants. For two Maxwellians at a common temperature this reduces to a one-dimensional integral over the relative speed $u$,

\begin{equation}
\begin{split}
    \langle\sigma v\rangle_M (T_M) = 4\pi &\left(\frac{\mu}{2\pi T_M}\right)^{\frac{3}{2}} \\
    &\times \int_0^\infty \sigma(u) u^3 \exp\left(-\frac{\mu u^2}{2T_M}\right) du
\end{split}
\end{equation}

\par The Thermal reactivity is computed in \cref{Thermal}.b again from the Bosch and Hale cross section parametrisation \cite{Bosch_and_Hale}.

\section{Non-thermal fusion reaction} \label{sec:Non-thermal fusion reaction}

\par As seen in \cref{Thermal}.a-b, the cross section (and then the reactivity as the cross section averaged over a velocity distribution) is an increasing functions of the relative energy of the colliding pair below the cross-section peak.
Accordingly, Nicolas (2026) \cite{Nicolas} came up with the idea of starting with a Maxwellian distribution, which he modified by accelerating a proportion $\alpha$ of ions particles to  high energy (where reactivity is important) represented by an isotropic Dirac distribution normalized to unity, leaving a proportion $(1-\alpha)$ in the Maxwellian distribution. (see \cref{fig:velocity_distribution}) In other words, this distribution of velocity consists of a Maxwellian distribution bulk combined with a hot tail, a high-energy Dirac distribution.

\begin{center}
    \captionsetup{type=figure}
    \nopagebreak
    \resizebox{\linewidth}{!}{%
    \begin{tikzpicture}[scale=0.9, >=latex]
        \begin{scope}[shift={(0,0)}]
            \draw[->, thick] (-0.5, 0) -- (5.5, 0) node[right] {$v_{th}$};
            \draw[->, thick] (0, -0.2) -- (0, 3.8);
            
            \node at (2.5, 4.3) {$f(v) = 4 \pi \left[\frac{m}{2\pi T_M}\right]^{3/2}v^2 \exp\left(-\frac{mv^2}{2T_M}\right)$};
            
            \draw[thick, domain=0:5, samples=100] plot (\x, {3 * exp(-(\x - 2)^2 / 0.7)});
            
            \draw[dashed, thick] (2, 0) -- (2, 3);
            \node[below] at (2, 0) {$v_{th} = \sqrt{\frac{2T_M}{m}}$};
        \end{scope}
        
        \node at (6.25, 1.8) {\Huge $+$};

        \begin{scope}[shift={(7.5,0)}]
            \draw[->, thick] (-0.5, 0) -- (5.5, 0) node[right] {$v$};
            \draw[->, thick] (0, -0.2) -- (0, 3.8);

            \node at (1, 4.3) {$f(v) = \frac{\delta(v-v_0)}{4\pi v_0^2}$};
            
            \draw[thick] (3.5, 0) -- (3.5, 2.5);
            
            \node[below] at (3.5, 0) {$v_0$};
        \end{scope}
    \end{tikzpicture}%
    }
    \caption{Velocity distribution model combining a thermal Maxwellian with $v_{th}$ the most probable velocity of this distribution and a hot tail non-thermal component with a unique velocity $v_0$.}
    \label{fig:velocity_distribution}
\end{center}

Therefore, the non-thermal isotropic distribution of velocity of a unique species can be expressed by:
\begin{equation}
    f(\vec{v},m,T_{M})=(1-\alpha) \underbrace{ \frac{\exp\left(-v^{2}/v_{th}^{2} \right)}{\pi^{3/2}v_{th}^{3}}}_{\text{Maxwell}}+\alpha  \underbrace{\frac{\delta(v-v_{0})}{4\pi v_{0}^{2}}}_{\text{hot tail}}
\end{equation}

$v_0$, is the speed of the particles in the hot tail (its choice is discussed below), and $\alpha$ is defined by:

\begin{equation}
    \alpha=\frac{\text{number of particles into the hot tail}}{\text{number of particles in total}}
\end{equation}

The total energy (in term of thermal energy) in this distribution:

\begin{equation}
    \mathcal{E}_{tot}=\frac{2}{3}\int \frac{1}{2} mv^{2}f(v)d^{3}v=(1-\alpha)T_{M}+ \underbrace{\frac{2\alpha}{3}E_{0}}_{\mathcal{E}_{tail}} \label{Etot_partial}
\end{equation}

with:

\begin{equation}
    E_{0}=\frac{1}{2}mv^{2}_{0}
\end{equation}

Defining $F$ as the fraction of energy in the hot tail, $\alpha$ can be expressed as a function of $F$:
\begin{align}
    &F=\frac{\mathcal{E_{\text{tail}}}}{\mathcal{E}_{tot}}=\frac{2\alpha E_{0}/3}{(1-\alpha)T_{M}+2\alpha E_{0}/3} \\
    &\leftrightarrow \alpha=\frac{F}{F+\frac{2E_{0}}{3T_{M}}(1-F)} \label{alpha}
\end{align}

\par The quantities $F$ and $\alpha$, measure the same thing, namely the size of the hot tail: $F$ does so in terms of energy, and $\alpha$ in terms of particle number. The relationship $\alpha(F)$ given by \cref{alpha} is non-linear; consequently a substantial energy fraction in the tail requires only a small fraction of the total particle population, because a single high-energy particle in the tail carries the energy of many thermal particles.

\par Throughout, reactant 1 is deuterium and reactant 2 may be tritium, helium-3, or deuterium. With this non-thermal distribution, four distinct particle-interactions are possible:

\begin{enumerate}
    \item Maxwellian–Maxwellian: both partners belong to the Maxwellian bulk
    \item hot tail (species 1)–Maxwellian (species 2)
    \item Maxwellian (species 1)–hot tail (species 2)
    \item hot tail–hot tail: both partners belong to the tail
\end{enumerate}

Define $\alpha_1$ and $\alpha_2$ as the respective hot-tail particle fractions for reactants 1 and 2, and assume $T_{M_1} = T_{M_2}$; All  four interactions then enter the reactivity of the non-thermal reaction:

\begin{equation}
    \langle \sigma v \rangle =\int_{0}^{\infty}\sigma(u)uh(u)du \qquad \int_{0}^{\infty}h(u)du=1
\end{equation}

where $h$ takes into account all  four interactions as a sum of four contributions:

\begin{align}
    h(u)=&(1-\alpha_{1})(1-\alpha_{2})h_{MM}(u) \\ \notag
    &+\alpha_{1}(1-\alpha_{2})h_{hM,12}(u) \\ \notag
    &+(1-\alpha_{1})\alpha_{2}h_{hM,21}(u) \\ \notag
    &+\alpha_{1}\alpha_{2}h_{hh}(u) \notag
\end{align}

and where the $h_{ij}$,  are the relative-speed distribution built from each pair of single-species distribution; each is independently normalized to unity.

\begin{equation}
    h_{\text{MM}}(u) = \frac{4}{\sqrt{\pi}} \left( \frac{\mu}{2T_M} \right)^{3/2} u^2 \exp\left( -\frac{\mu u^2}{2T_M} \right) \label{eq:18}
\end{equation}

\begin{equation}
    h_{\text{hM},12}(u) = \frac{2}{\sqrt{\pi}} \frac{u}{v_0 v_{\text{th},2}} \exp\left( -\frac{u^2 + v_0^2}{v_{\text{th},2}^2} \right) \sinh\left( \frac{2uv_0}{v_{\text{th},2}^2} \right) \label{eq:19}
\end{equation}

\begin{equation}
    h_{\text{hM},21}(u) = \frac{2}{\sqrt{\pi}} \frac{u}{v_0 v_{\text{th},1}} \exp\left( -\frac{u^2 + v_0^2}{v_{\text{th},1}^2} \right) \sinh\left( \frac{2uv_0}{v_{\text{th},1}^2} \right) \label{eq:20}
\end{equation}

\begin{equation}
    h_{\text{hh}}(u) = \begin{cases} 
        \dfrac{u}{2v_0^2} & \text{if } u < 2v_0, \\[6pt]
        0 & \text{otherwise.} 
    \end{cases} \label{eq:21}
\end{equation}

\begin{figure*}[htbp]
    \centering
    \includegraphics[width=1\linewidth]{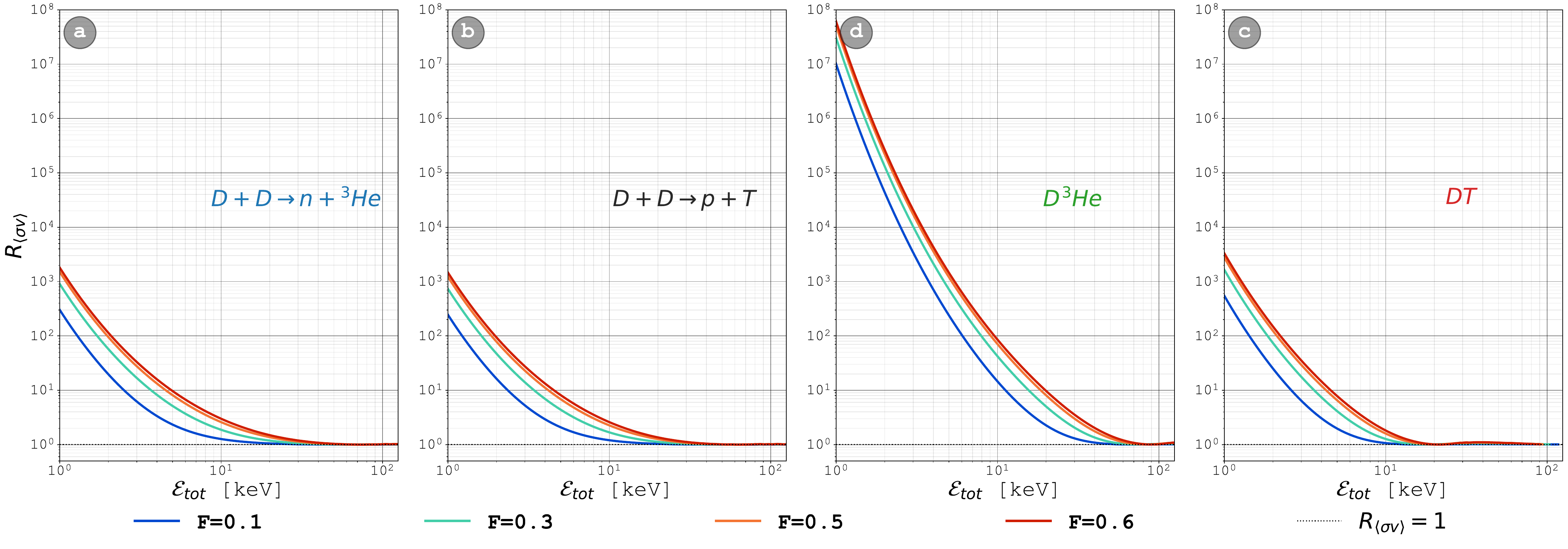}
    \caption{Reactivity increase rate $R_{\langle \sigma v \rangle}$ as a function of total energy for (a) $D+D \to n+ \ce{^3He}$ reaction, (b) $D+D \to p +T$ reaction,(c,) $D-T$ reaction and (d) $\ce{D-^3He}$ reactions} 
    \label{Rrate}
\end{figure*}

\par These four distributions are derived in Ref.~\cite{Nicolas}; they are quoted here only to fix the notation used in the rest of this paper.

\par Setting the two species to the same energy fraction, $F_1=F_2\equiv F$, the combined distribution reduces to $h(u,T_M,v_0,F)$. Equal energy fractions do not imply equal particle fractions: $\alpha$ depends on $E_{0i}=\tfrac{1}{2}m_iv_0^2$, so at fixed $F$ the lighter reactant always has the larger tail population, $\alpha_1(F,v_0)>\alpha_2(F,v_0)$ for $F\in\,]0,1[$.

Let's now define a new total energy for this new distribution, assuming a stoichiometric fuel ($n_1=n_2)$:

\begin{equation}
\begin{split}
\mathcal{E}_{\text{tot}}(T_M,F,v_0) &= \frac{1}{2} \bigg[ (1-\alpha_{1})T_{M}+\frac{2\alpha_{1}}{3}E_{01} \\
&\quad + (1-\alpha_{2})T_{M}+\frac{2\alpha_{2}}{3}E_{02} \bigg] \label{Etot}
\end{split}
\end{equation}

\par The  goal is to maximize the fusion reactivity relative to a purely thermal baseline at  constant total energy $\mathcal{E}_{\mathrm{tot}}$. The reactivity depends on three parameters, $\langle \sigma v \rangle(T_M, F, v_0)$. Let’s consider $v_0$ as a parameter of our system and determine the optimal velocity that should be parametrized to our hot tail in order to achieve the highest reactivity while maintaining a fixed total energy :

\begin{equation}
    v_{opt}(T_{M})=\underset{v_{0}}{arg \ max} \ \frac{\partial \langle\sigma v \rangle}{\partial F}_{|\mathcal{E}_{tot}=C^{te},F=0}(T_{M})
\end{equation}

The constrained derivative is expanded in Ref.~\cite{Nicolas}.

\begin{center}
    \centering
    \captionsetup{type=table}
    \caption{Peak of energy in term of energy of center of mass $E_{c.m}$ [keV] then in term of deuterium energy $E_D$ [keV] for D-D, D-T, and $\ce{D-^3He}$ reactions from \cite{E_peak} [table 1.1, p.14] }
    \label{table Ed}
    
    \begin{tabular}{lcr}
        \toprule
        \textbf{Reaction} & \textbf{$E_{c.m}$ peak} & \textbf{$E_D$ peak } \\ 
        \midrule
        $D-D \to \ce{^3He}+n$ & 1750 keV & 3500 keV  \\ 
        $D-D \to T + p$ & 1250 keV & 2500 keV \\ 
        $\ce{D-^3He}$ & 250 keV & 417 keV \\
        $D-T$ & 64 keV & 107 keV \\
        \bottomrule
    \end{tabular}
\end{center}

\cref{Optimal tail energy} reproduces $E_{D,\text{opt}}(T_M)=\tfrac{1}{2}m_Dv_{opt}^2$ from Ref.~\cite{Nicolas}. Below $7~\text{keV}$ the optimum for $\ce{D-^3He}$ and D--T sits close to the cross-section peak (\cref{table Ed}); for the D--D branches, whose peaks lie at 2500 and $3500~\text{keV}$, the energy cost of the tail forces a compromise well below the peak.

\begin{center}
    \captionsetup{type=figure, justification=raggedright}
    \centering
    \includegraphics[width=1\linewidth]{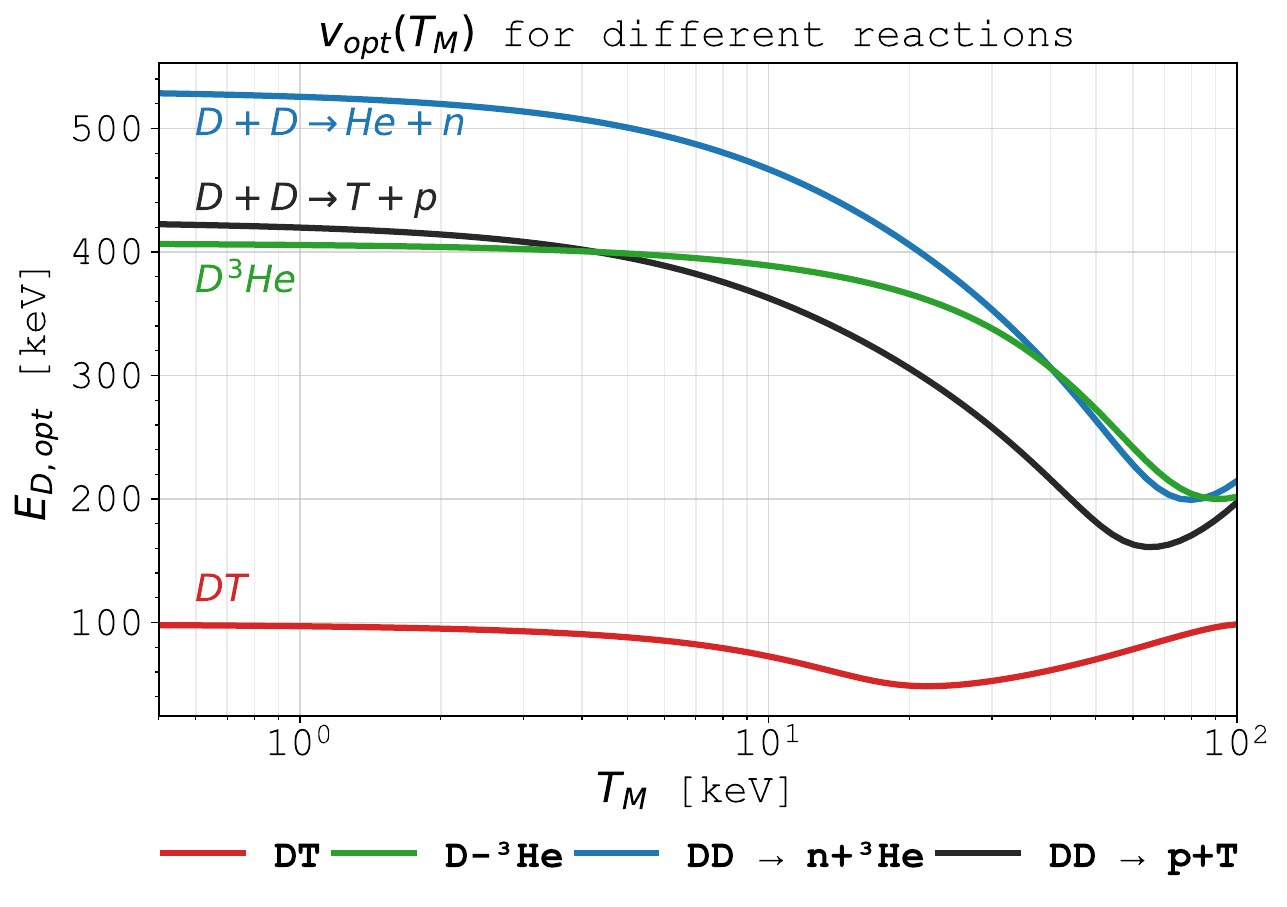}
    \caption{Optimal deuterium energy, $E_{D,\text{opt}}(T_M) = \frac{1}{2} m_D v_{\text{opt}}^2(T_M)$, for D-T (red), D-$^3$He (green) and D-D (blue and black) reactions as a function of $T_M$. Recomputed from \cite{Nicolas} [Figure.1, p.5] which validate the authors' original findings}
    \label{Optimal tail energy}
\end{center}

Let's define a new parameter in order to quantify this enhancement at constant energy. A new ratio has been defined, the reactivity increase rate as:
\begin{equation}
    R_{\langle \sigma v \rangle}(\mathcal{E}_{tot},F)= \frac{\langle \sigma v \rangle(\mathcal{E}_{tot},F)}{\langle\sigma v \rangle_{M}(\mathcal{E}_{tot})}
\end{equation}

\begin{figure*}[htbp]
    \centering
    \includegraphics[width=1\linewidth]{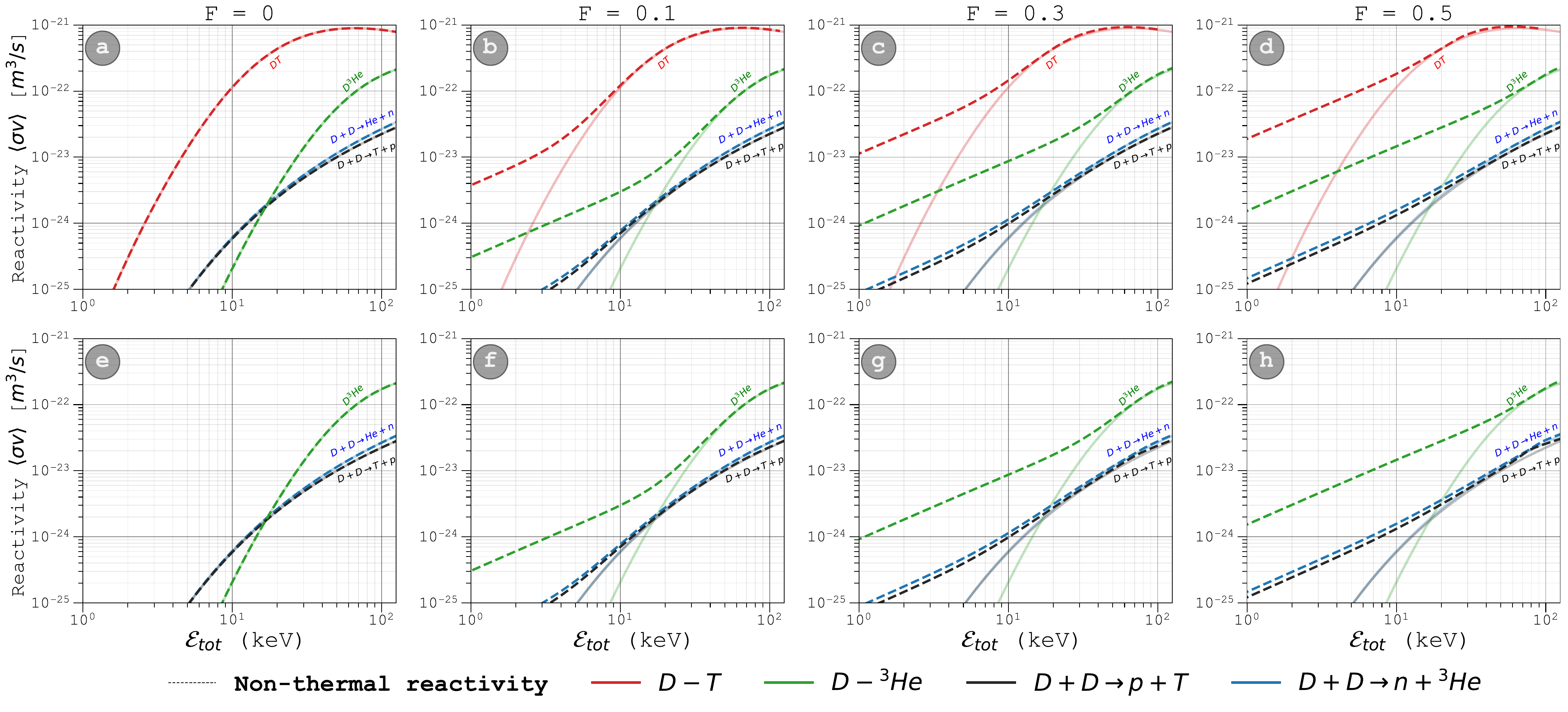}
    \caption{Comparison of non-thermal (dashed lines) and thermal (transparent curves) reactivities for different values of $F$ in D-D, D-T, and D-$^3$He reactions (top (a),(b),(c) and (d)) and into $\ce{D^3He}$ fuel where $v_{opt}(T_M,r)=v_{opt}(T_M,r=\ce{D-^3He})$ (bottom (e),(f),(g),(h)) } 
    \label{NT Reactivity}
\end{figure*}

\par This equation tell us, at a specific total energy $\mathcal{E}_{tot}$ what is the value of that gain relative to a Maxwellian holding the same energy. The reactivity increase rate $R_{\langle \sigma v \rangle}$ is plotted in \cref{Rrate} for D-T, $\ce{D-^3He}$ and D-D reactions. In these plots, overall $R_{\langle \sigma v \rangle}$ is a decreasing function of $\mathcal{E}_{tot}$ and an increasing function of F; and for every
$F \in [0, 0.6]$ there is a total energy within [1, 200] keV above which $R_{\langle \sigma v \rangle} \approx 1$. However $R_{\langle \sigma v \rangle}$ at low energy reaches about three orders of magnitude for D--D and D--T, and between seven and eight orders of magnitude for $\ce{D-^3He}$. The $\ce{D-^3He}$ reaction seems to be the most promising reaction like what Nicolas had found \cite{Nicolas}, because, it has the highest increase rate overall: it starts from about seven to eight orders of magnitude at the lowest energies and still gives roughly one order of magnitude at 10 keV for $F=0.1$ and two orders of magnitude for $F=0.6$ whereas for the other reactions the gain at $10~\text{keV}$ is at most a factor of about 3

\par Nicolas \cite{Nicolas} cautions that this $\ce{D-^3He}$ gain is not accessible at 10-20 keV, because the Maxwellian minimum ideal ignition energy is around 30 keV, and reaching this high-enhancement regime would require the electron temperature to be reduced. However, we will see in \cref{sec:Ignition energy}  that this is not necessary.

\par In a real $\ce{D-^3He}$ fuel all three reactions share one hot tail, so the tail velocity cannot be optimised for each separately; we therefore set $v_0=v_{opt}(T_M,\ce{D-^3He})$ for all three. This makes no visible difference: the D--D reactivities computed with their own optima or with the $\ce{D-^3He}$ optimum agree to within the line width, because the two optima are close and follow the same trend (\cref{Optimal tail energy}) and their reactivity enhancements are low (\cref{Rrate}).

\par Then \cref{NT Reactivity} is the non-thermal counterpart of \cref{Thermal}.b, for $F=0.0$, $0.1$, $0.3$ and $0.5$. The $\ce{D-^3He}$ non-thermal reactivity is the most impressive one, as \cref{Rrate} already showed. The $\ce{D-^3He}$ curve, is lifted above both D--D branches across the whole $1$--$100~\text{keV}$ range.

Spacing function is defined by the quotient of the non-thermal $\ce{D-^3He}$ reactivity by the non-thermal $D-D$ neutron branch reactivity:
\begin{equation}
    \mathcal{S}(\mathcal{E}_{tot},F)=\frac{\langle \sigma v \rangle_{\ce{D-^3He}}(\mathcal{E}_{tot},F)}{\langle \sigma v \rangle_{D-D}(\mathcal{E}_{tot},F)}
\end{equation}

\par The spacing function is plotted in \cref{spacing}: $\mathcal{S}$ is an increasing function of $F$, so the larger $F$ is, the wider the separation. Even an energy fraction of $F=0.1$ keeps $\mathcal{S}>1$ for every value of total energy between $1$ and $100~\text{keV}$. Furthermore, every curve with $F \geq 0.1$ falls to a minimum before rising again; and above about 80 keV all seven curves converge on $\mathcal{S} \simeq 6.5$, so the ordering by F no longer holds there. At $F=0$, the curve crosses $\mathcal{S}=1$ at $\simeq 17$ keV which matches the thermal crossover in \cref{Thermal}.b
 
\par As shown by this spacing plot \cref{spacing}, the greater the reactivity gap is (large spacing), more probable would be the primary reaction ($\ce{D-^3He}$) compared to the secondary reactions (D-D) and their corresponding reaction products. As a result, if a specific output power is set, our non-thermal regime would produce fewer neutrons than in the thermal case, and increasing the proportion of energy in the hot tail F would only accentuate this phenomenon.

\begin{center}
    \captionsetup{type=figure, justification=raggedright}
    \centering
    \includegraphics[width=1\linewidth]{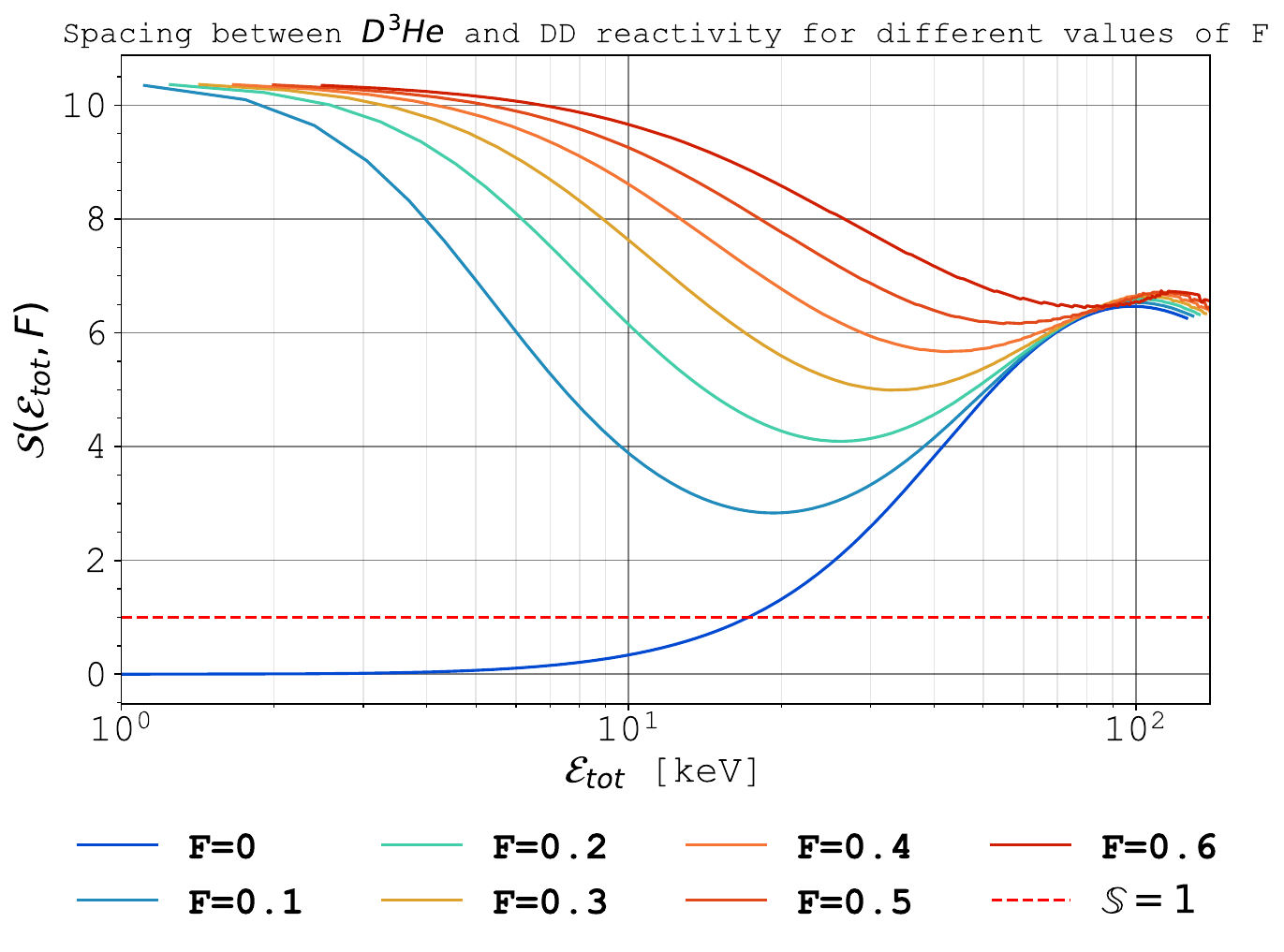}
    \caption{Spacing between \ce{D-^3He} and $D+D \to n+ ^3He$ non-thermal reactivity for different values of $F$ into $D-^3He$ fuel}
    \label{spacing}
\end{center}

\begin{figure*}[hbtp]
    \centering
    \includegraphics[width=1\linewidth]{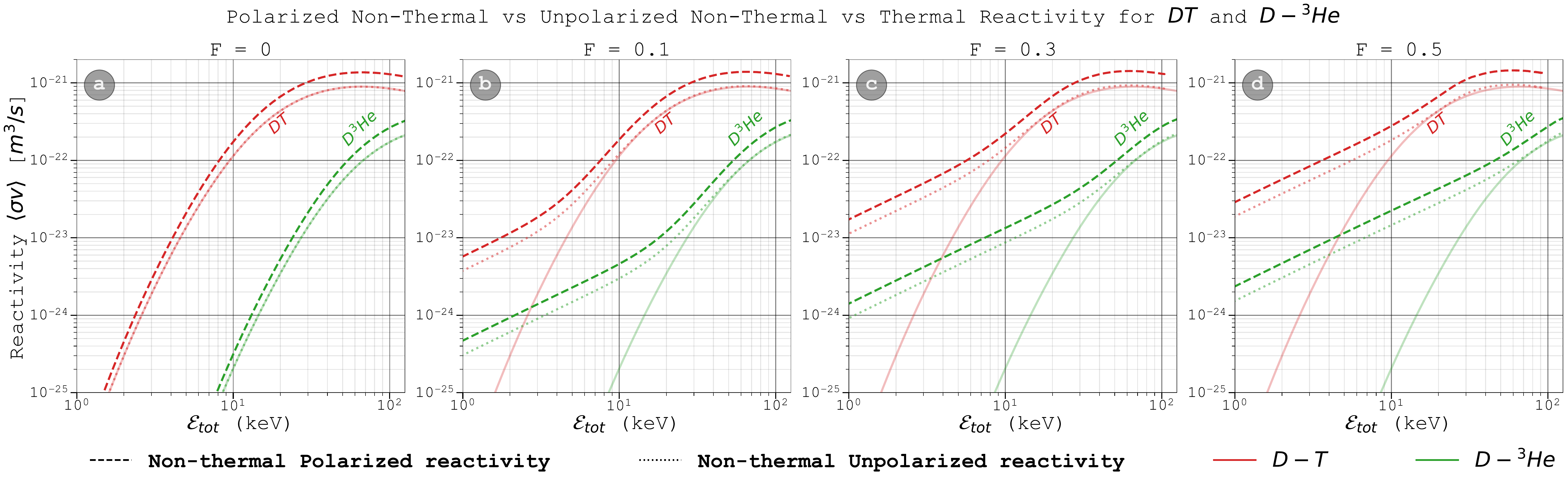}
    \caption{Comparison of polarized non-thermal (dashed lines), non-thermal (transparent dotted line) and thermal (transparent curves) reactivities for different values of F for D-T, and D-3He reactions} 
    \label{Polarized Reactivity 3N}
\end{figure*}

\par We have seen in this section that the non-thermal regime, consisting of a Maxwellian bulk and a hot tail, increases the reactivity by several orders of magnitude at low energy ($\ll 10~\text{keV}$), but has little or no effect at higher energy ($\ge 10~\text{keV}$). The deuterium--helium-3 reaction is the most promising of those considered, as it exhibits the largest increase in reactivity and the widest energy range over which the enhancement persists before the effect diminishes. We have also seen that, in a deuterium--helium-3 fuel, the non-thermal distribution opens a gap in reactivity between the primary reaction \cref{DHe3} and the secondary deuterium--deuterium reactions \cref{ddn,ddp}, by a factor that increases with the energy fraction in the hot tail.

\section{Polarized non-thermal fusion reaction} \label{sec:Polarized non-thermal fusion reaction}

\begin{figure*}[hbtp]
    \centering
    \includegraphics[width=1\linewidth]{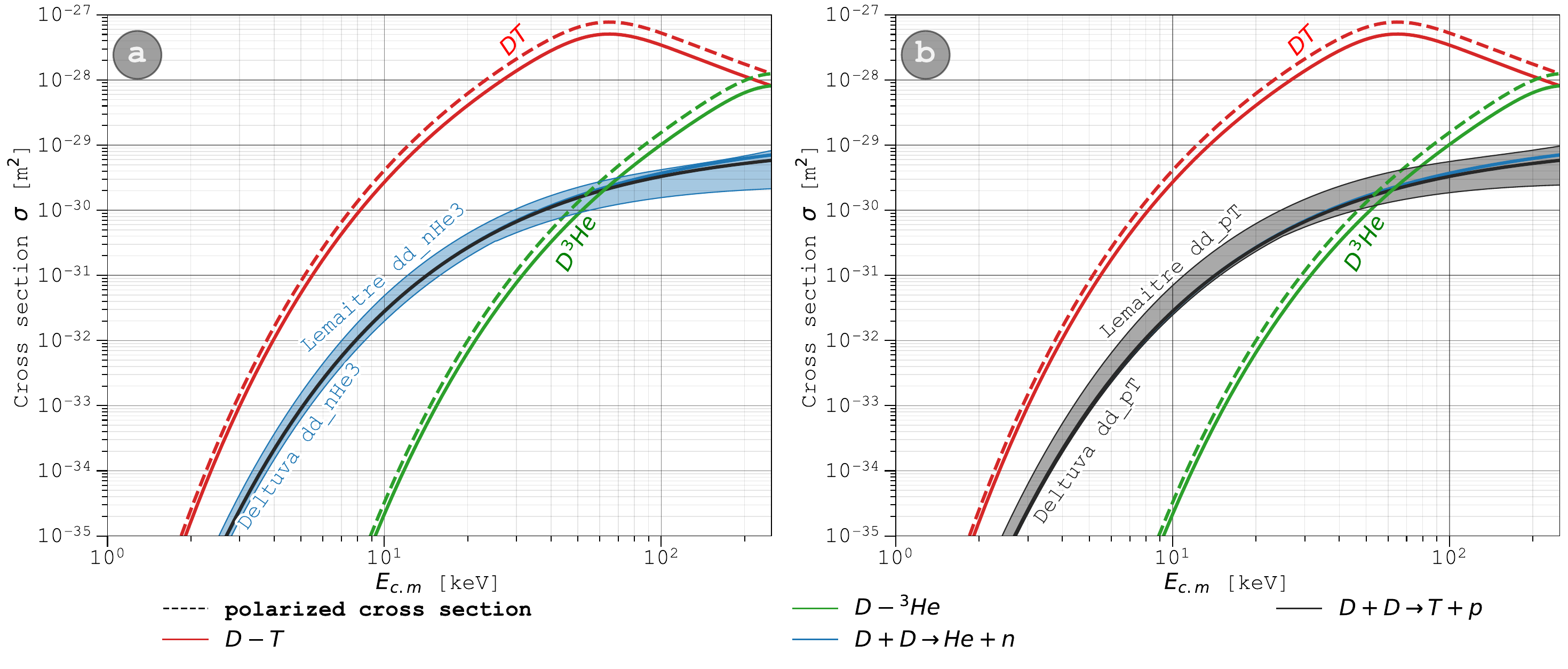}
    \caption{Cross section comparison for D-T, $\ce{D-^3He}$ and D-D reactions. For D-T and $\ce{D-^3He}$, unpolarized (solid lines) and polarized (dashed). Lemaître's evaluation of QSF represents the upper bound for QSF, while Deltuva's evaluation of QSF represents the lower bound. Panel (a), left, focuses on the D--D neutron branch 2 \cref{ddn} and panel (b), right, on the D--D proton branch 1 \cref{ddp}.} 
    \label{Polarized Cross Section}
\end{figure*}

\par The $\ce{D-^3He}$ and D--T reactions proceed through the intermediate compound states $\ce{^5Li}$ and $\ce{^5He}$ respectively, which then decay into the reaction products. These states have a total angular momentum $J^{\pi}=3/2^+$ \cite{Kulsrud, DD_from_HD_target}, so the state is strongly favored when the incident ions couple to a total angular momentum of $J=3/2$. Here, $J$ is given by $\vec{J}=\vec{\ell}+\vec{S}$, where $\vec{\ell}$ is the orbital angular momentum and $\vec{S}$ is the total spin angular momentum; however, at low energies on the order of the temperatures found in fusion devices, $\vec{\ell}=\vec{0}$ ($s$-wave). Thus, the total spin $S$ must equal $3/2$, since the deuteron has spin $s_D=1$ (and is therefore a boson) while tritium and helium-3 have spin $s_{3N}=1/2$. (their respective single nuclear spin state projection along z-axis are $m_i \in \{ x\in[-s_i,s_i] \ | x-s_i \in \mathbb{Z}\}$ so $m_{D}=(1,0,-1)$ and $m_{3N}=(1/2,-1/2)$). so their possible combined spin states ($S \in \{ x \in [|s_D-s_{3N}|,s_D+s_{3N}] \ | x-|s_D-s_{3N}|\in \mathbb{Z} \}$) gives $S=3/2 $ or $S=1/2$. 

\par In an unpolarized fuel, the nuclear spins are randomly distributed so 6 states are equally likely ($m_s=\{a+b \ | \ a\in m_D,b\in m_{3N}\}, |m_s|=|m_D| \times |m_{3N}|=6$) so that, statistically, there are 4 spin-states combinations which leads to a total spin of 3/2 ($m_s=(3/2, 1/2, -1/2, -3/2)$) and 2 spins-states combinations which leads to a total spin of 1/2 ($m_s=(1/2,-1/2)$). The principle of polarized fuel is to ensure that the only possible spin-state is the one which gives an increase of a factor of 1.5 in the cross section: when unpolarized, 6 spin states are accessible and only 4 of them carry the reaction, whereas with fully parallel spins the entire population sits in the $S=3/2$ channel, so the rate is multiplied by $6/4=1.5$. This was established by Kulsrud et al. \cite{Kulsrud, Kulsrud_Polarized_fusion_reactor} (1982 and 1986). 

\par Another effect of polarized fuel is that fusion product emission is no longer isotropic; this effect is not studied here. A recent paper by Parisi et al. \cite{parisiEnhancedAlphaChanneling2026} shows that the preferential emission of alpha particles perpendicular to the magnetic field \cite[Fig.~6]{parisiEnhancedAlphaChanneling2026} enhances the alpha-channeling effect, in which the alpha-particle energy is transferred directly to the fuel ions, driving a temperature separation between $T_i$ (ion temperature) and $T_e$ (electron temperature) \cite[Fig.~9]{parisiEnhancedAlphaChanneling2026}. This direct increase in $T_i$ yields an additional boost in reactivity, resulting in a non-linear enhancement relative to the bare polarization gain. In a zero-dimensional ARC-class model, the power enhancement reaches a factor of 4.7. In an older study of a reactor with dimensions similar to those of ITER, \cite{in_situ_tests} this gain was estimated to be closer to 1.8 for fully polarized ions.

The differential cross section, denoted $\frac{d \sigma}{ d\Omega}$, in its most general sense, measures the number of particles scattered into a specified final-state region, normalized by the incident flux and the solid angle $d \Omega=\sin(\theta) d \theta  d \phi$. For a reaction between a deuterium and a 3 nucleons-nucleus (Tritium or helium-3), in which both reactants are colliding in a single direction along the z-axis, the polarized differential cross section is simply given by \cite{Ab_initio_DT}:

\begin{equation}
\begin{split}
    \frac{\partial \sigma_{pol}}{\partial \Omega_{c.m.}}=  &\frac{\partial \sigma_{unpol}}{\partial \Omega_{c.m.}}(\theta_{c.m.}) \\
    &\times \left( 1+\frac{1}{2}p_{zz}A_{zz}^{(b)} (\theta_{c.m})+\frac{3}{2}p_{z}q_{z}C_{z,z}(\theta_{c.m.}) \right) \label{brut_differential_cross_section}
\end{split}
\end{equation}

where $A_{zz}^{(b)}$ is the beam tensor analyzing power and $C_{z,z}$ is the spin correlation coefficient. $q_z$ is the spin projection along the $z$-axis of the three-nucleon nucleus, and $p_z$ is the spin projection along the $z$-axis of the deuteron, with $q_z = \mathbb{P}(1/2) - \mathbb{P}(-1/2)$ and $p_z = \mathbb{P}(1) - \mathbb{P}(-1)$, such that $(p_z, q_z) \in [-1, 1]^2$. The quantity $p_{zz}$ is the tensor polarization of the deuteron along the $z$-axis, which measures the depletion of the $m_D = 0$ population relative to its unpolarized, isotropic value:
$p_{zz} = \mathbb{P}(1) + \mathbb{P}(-1) - 2\mathbb{P}(0) = 1 - 3\mathbb{P}(0)$. 
\par $\mathbb{P}(m_i)$ is a probability of being in a nuclear state projection $m_i$ where $m_i=m_D$ for the deuteron and $m_i=m_{3N}$ for the three-nucleon nucleus \cite{in_situ_tests}.
This equation is a simplification of the general polarized differential cross section \cite[Appendix, Eq.~2.14]{Where_do_we_stand}, valid when the two reactants collide along a single axis; in a tokamak, this corresponds to the beam-target geometry set by the toroidal magnetic field.

As mentioned earlier, at low energy the orbital angular momentum $\ell$ is zero (s-wave) and it impacts the previous equation by the fact that $A_{zz}(\theta_{c.m})$ and $C_{z,z}(\theta_{c.m})$ can be expressed using Legendre polynomials \cite{Ab_initio_DT} and \cref{brut_differential_cross_section} can be simplified into \cite{in_situ_tests} [eq.4a]:

\begin{equation}
\begin{split}
    \frac{d\sigma_{\text{pol}}}{d\Omega} &= \frac{d\sigma_{\text{unpol}}}{d\Omega} \left( 1 - \frac{1}{2} p_z q_z \right. \\
    & \left. + \frac{1}{2} \left[ 3 p_z q_z \sin^2(\theta_{\text{c.m.}}) + \frac{1}{2} p_{zz} \left( 1 - 3 \cos^2(\theta_{\text{c.m.}}) \right) \right] \right)
\end{split}
\end{equation}

When we integrate this equation over $\Omega$, we get:

\begin{equation}
    \sigma_{pola}= g_{pol}(p_z,q_z) \sigma_{unpola}, \quad g_{pol}(p_z,q_z)= 1+ \frac{p_zq_z}{2} \label{eq:gpola}
\end{equation}

Clearly, when the plasma is fully polarized in one direction, that is when $p_zq_z=1$,we get $\sigma_{pola}=1.5 \times \sigma_{unpola}$ which is exactly what we found above and what has been theorized by Kulsrud \cite{Kulsrud}. The map $(p_z,q_z) \to g_{pol}(p_z,q_z)$, is plotted in \cite[Figure.2]{Parisi_DT}. 

\par However, in the context of our study, it would not be correct to assume $\ell$ = 0 because our distribution is non-thermal and contains high-energy particles exceeding several keV (see \cref{Optimal tail energy}).
\par Fortunately, in 2019, calculations have been performed on D--T reaction using the ab initio method, which made it possible to predict the behaviour of the coefficients $A_{zz}$ and $C_{z,z}$ at higher energy including the $\ell=2$ (d-wave) contribution. However, it appears that $\underset{p_z,q_z,p_{zz}}{\max} g_{pola}$ increases, reaching a maximum around 200 keV, and then gradually decrease at higher energies, reaching a value of 1.4 around 900 keV \cite[Fig.4a]{Ab_initio_DT}. Since the D--He3 reaction is essentially the mirror image of the D--T reaction, we expect that $g_{pola}$ for the $\ce{D-^3He}$ reaction will behave like that of the D--T reaction, but with its maximum shifted to a higher energy \cite{discussion_DT}.

\par It was noted that, even when contributions up to the d-wave ($\ell$=2) were taken into account, $g_{pola}$ was not enough quantitatively different from the case where only the s-wave was considered (see Appendix \labelcref{sec:constant gain}). Consequently, without any loss of accuracy, it is more than sufficient to use $g_{pola}$ given by \cref{eq:gpola}

\par The polarized non-thermal reactivity is plotted for D--T and $\ce{D-^3He}$ reactions for different values of $F$ in \cref{Polarized Reactivity 3N}.

\par The equations for polarized non-thermal D–T and $\ce{D-^3He}$ reactions are given by:
\begin{align}
\langle \sigma v \rangle_{pola}(v_{0})&= \int_{0}^{\infty} \sigma_{pola}(u)\,u\,h(u,T_{M},F,v_{0})\,du \\
&=g_{pola} \langle \sigma v \rangle_{unpol}(v_{0})
\end{align}

Now let's determine the optimal velocity for a polarized plasma:
\begin{align}
v_{opt,pola}(T_{M})&=\underset{v_{0}}{arg \ max} \ \frac{\partial \langle\sigma v \rangle_{pola}}{\partial F}_{|\mathcal{E}_{tot}=C^{te},F=0}(T_{M}) \\
&=v_{opt,unpol}(T_{M})
\end{align}

\par Since the optimal speed is not affected by polarization, our new reactivity is simply increased by a factor of $g_{pola}$; we will therefore assume a fully polarized plasma, so $g_{pola}=1.5$ :

\begin{equation}
\langle \sigma v \rangle_{pola}(v_{opt,pola})=g_{pola}\langle\sigma v \rangle_{unpol}(v_{opt,unpol})
\end{equation}

\par While this describes the polarization of D--T and $\ce{D-^3He}$ reactions, the situation is considerably more complex for the two D--D reactions. The deuteron has spin $s_D = 1$ (hence it is a boson), and its projection along a quantization axis takes the values $m_D \in \{-1, 0, 1\}$; two spin-1 particles therefore admit many more spin combinations than the spin-1 / spin-$1/2$ systems discussed above. Consequently, the general spin-1 on spin-1 differential cross section expression is far more intricate \cite[Appendix, Eq.~2.15]{Where_do_we_stand}. Analogous to the parameter $g_{pola}$ for D--T and $\ce{D-^3He}$, when two polarized deuterons (both with $m_D = 1$) interact---that is, in the ``quintet'' $S=2$ channel---the Quintet Suppression Factor (QSF) is defined as the ratio of the polarized D--D cross section $\sigma_{1,1} = \sigma_{\text{pol}}$ to the unpolarized cross section.

\begin{center}
    \captionsetup{type=figure}
    \centering
    \includegraphics[width=1\linewidth]{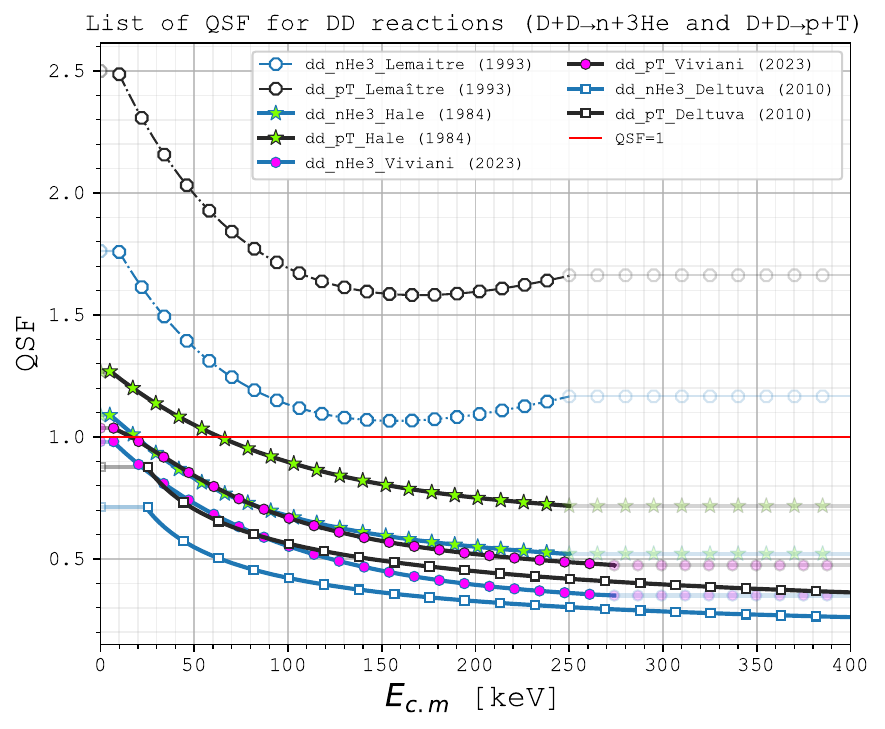}
    \caption{Theoretical QSF values: Hale and Doolen (1984) \cite{Doolen},  Lemaître et al. (1993) \cite{Lemaitre},  Deltuva et al. (2010) \cite{Deltuva} and Viviani et al. (2023) \cite{Viviani} . Adapted from Ref. \cite[Fig.2, p. 3]{Viviani}. The curves for Hale and Doolen are fitted between 5 and 250 keV, those for Lemaître between 10 and 250 keV and those for Viviani er al. between 7 keV and 274 keV; outside this range, they are held constant at their respective boundary values (transparent curve).The same approach was applied to Deltuva et al. over a wider range from 25 keV to 400 keV. dd\_nHe3: D--D neutron branch \cref{ddn} and dd\_pT: D--D proton branch \cref{ddp}}
    \label{all QSF}
\end{center}

\begin{equation}
    QSF=\frac{\sigma_{pola}}{\sigma_{unpol}}
\end{equation}

This factor has been called as a suppression factor because first calculation displayed that when deuterons were polarized, it could result in a "suppression" of $D+D \to n+ \ce{^3He}$ reaction which leads to imagine a neutron-free $\ce{D-^3He}$ reaction \cite{Kulsrud, Zhang_DD_first}. It's been a long time, since attempts have been made to theorize and evaluate the QSF, the first evaluation was that of Hale and Doolen (1984) using multi-channel $R$-matrix theory for the four-nucleon ($^4\text{He}$) system \cite{Doolen}. Then Stéphane Lemaître (in collaboration with H. Paetz gen. Schieck in 1990 and 1993) evaluated the behavior of the Quintet Suppression Factor (QSF) using a data-driven Transition Matrix Element (T-Matrix) analysis and he has concluded about the impossibility of the suppression of D-D reaction: \textit{"Over the entire energy region under discussion the quintet-correlation cross sections for both reactions remain enhanced or at least equal relative to the unpolarized cross sections $\sigma$. Neutron suppression will not take place in polarized deuterium fuel."} \cite{Lemaitre}. In 1999 Zhang, Liu and Shuy obtained a markedly different result: the QSF is about $0.86$ at $30~\text{keV}$ and falls to $0.22$ at $90~\text{keV}$ only for the proton-Tritium D--D branch, a strong suppression but no other publication has reproduced such an evaluation \cite[Table~V, Fig.~8]{Zhang_DD}. In 2010, Deltuva and Fonseca solved the exact four-particle equations in momentum space and found the cross section for spin-aligned deuterons to be significantly suppressed at several $\text{MeV}$, but not in the low-energy $\text{keV}$ regime relevant to thermonuclear fusion \cite[Fig.~9]{Deltuva}. Most recently in 2023, Viviani and collaborators evaluated the QSF using an \textit{ab initio} hyperspherical harmonics method; their results show close agreement, above roughly 150 keV, with the calculations of Deltuva and Fonseca (2010) \cite{Deltuva} and Hale and Doolen (1984) \cite{Doolen, Viviani}. However, some experimental data are still missing to confirm which approach describes physical reality; thus, all four D--D polarized QSF models will be considered and compared: Hale and Doolen (1984), Lemaître (1993), Deltuva and Fonseca (2010), and Viviani (2023). Zhang et al.\ (1999) is excluded here because its QSF evaluation spans too narrow an energy range and departs substantially from the other three. All QSF evaluations are plotted in \cref{all QSF}.

Cross section for $D-D$ reactions (see \cref{Polarized Cross Section}):
\begin{equation}
    \sigma_{pola}(E)=QSF(E) \sigma_{unpol}(E)
\end{equation}

Then reactivity:
\begin{equation}
\begin{split}
    \langle \sigma v \rangle_{pola}(T_M,F)=\int_{0}^{\infty}QSF(u)&\sigma_{unpol}(u)u  \\
    & \times h(u,T_M,F,v_{0,pola}) du
\end{split}
\end{equation}

\begin{figure*}[t]
    \centering
    \includegraphics[width=1\linewidth]{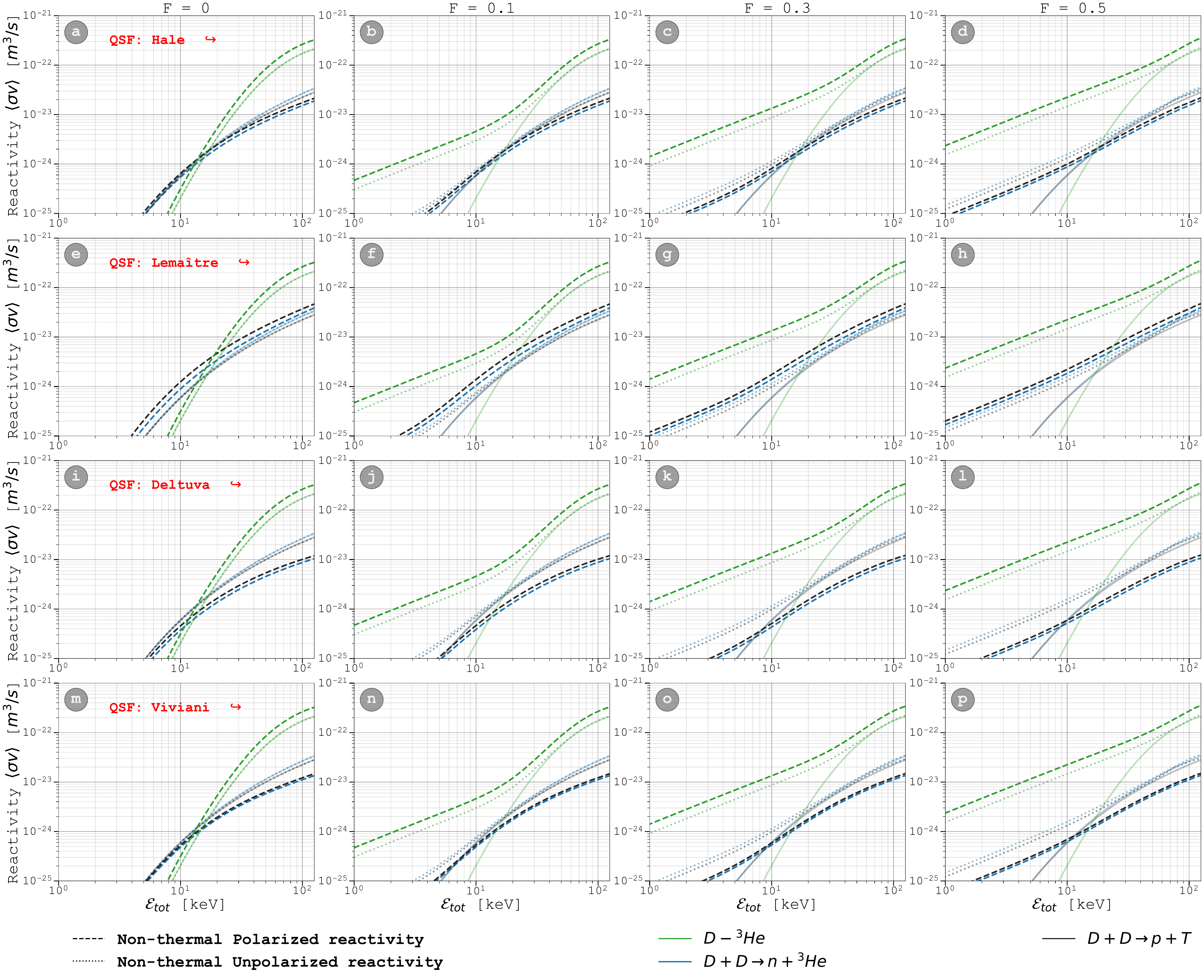}
    \caption{Comparison of polarized non-thermal (dashed lines), non-thermal (transparent dotted line) and thermal (transparent curves) reactivities for different values of F (each column) into Polarized $\ce{D-^3He}$ fuel for different QSF evaluation: Hale and Doolen (1984) \cite{Doolen} (a,b,c,d), Lemaître (1993) \cite{Lemaitre} (e,f,g,h), Deltuva (2010) \cite{Deltuva} (i,j,k,l) and Viviani (2023) \cite{Viviani} (m,n,o,p)}
    \label{Polarized_DHe3_Reactivity}
\end{figure*}

Because the goal here is not to optimize the D--D reaction but to examine the effect of its polarization on the $\ce{D-^3He}$ reaction, the hot-tail velocity is taken from the polarized $\ce{D-^3He}$ optimum for all three reactions. The resulting reactivities are shown in \cref{Polarized_DHe3_Reactivity}.

\par
\begin{figure*}[t]
    \centering
    \includegraphics[width=1\linewidth]{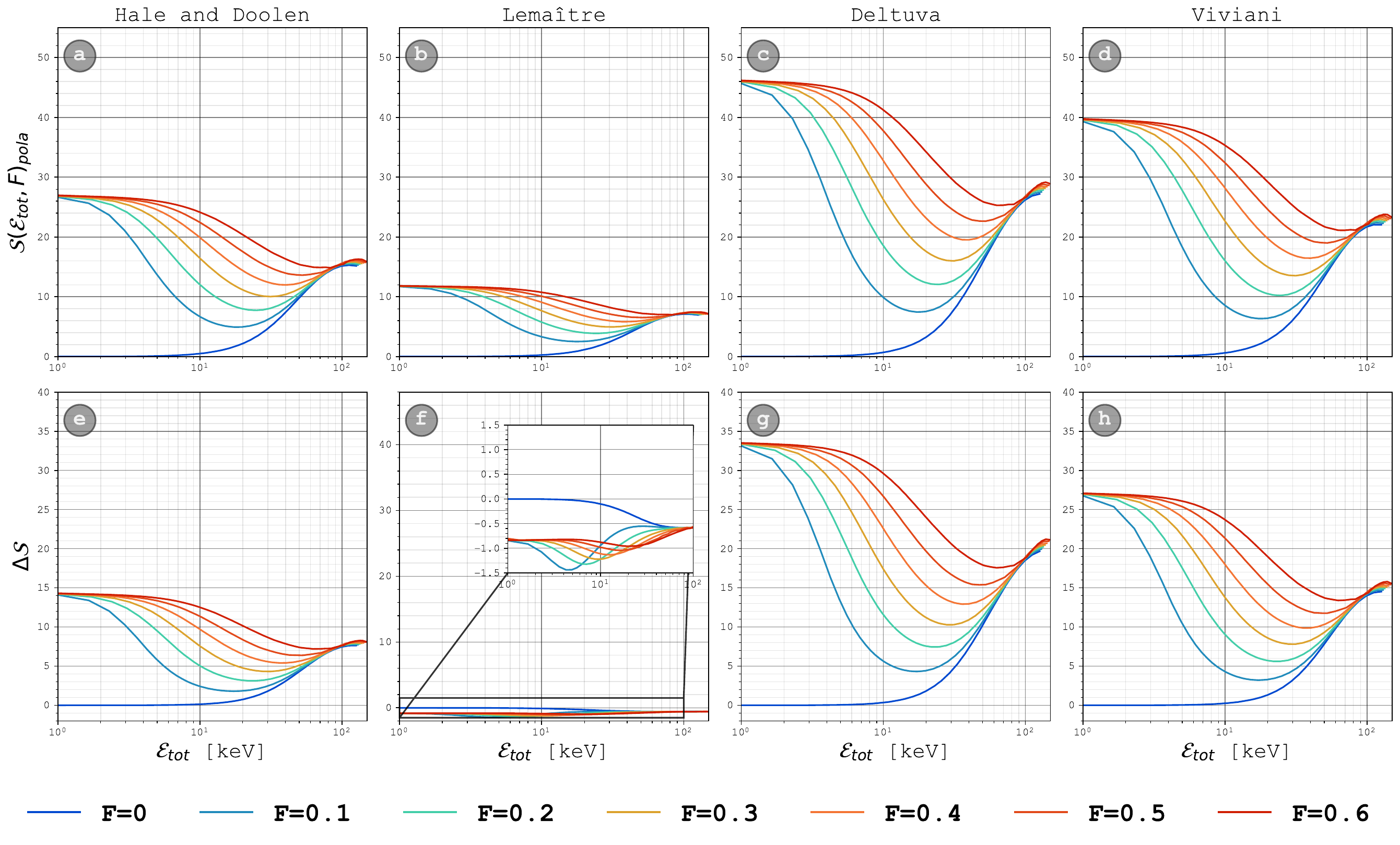}
    \caption{Spacing between $\ce{D-^3He}$ and $D+D \to \ce{^3He} +n$ polarized non-thermal reactivity for different values of F into $\ce{D-^3He}$ fuel for different QSF evaluation (top figures) and subtracted spacing from polarized non-thermal reactivity to unpolarized non-thermal reactivity for different values of F into $\ce{D-^3He}$ fuel for different QSF evaluation (bottom). Hale and Doolen (1984) \cite{Doolen} (a,e), Lemaître (1993) \cite{Lemaitre} (b,f), Deltuva (2010) \cite{Deltuva} (c,g) and Viviani (2023) \cite{Viviani} (d,h)}
    \label{Spacing}
\end{figure*}

 The four QSF evaluations are plainly different in \cref{all QSF} --- at $E_{c.m}\to0$ they span $\sim0.7$ to $\sim 2.5$, a factor of $3.5$ --- yet the four rows of \cref{Polarized_DHe3_Reactivity} look almost identical. This is a property of the plot, not of the physics. \cref{Polarized_DHe3_Reactivity} is logarithmic over four decades of reactivity, and the QSF multiplies only the two D--D curves: a factor of $3.5$ is half a decade, or about a tenth of the axis, and between neighbouring models (Deltuva $0.71$ against Viviani $1.05$) it is a factor of $1.5$, some $4\%$ of the axis. The choice of QSF is therefore invisible on this figure and visible on any derived quantity that isolates the D--D contribution. The spacing $\mathcal{S}_{pola}$ of \cref{Spacing} is such a quantity, and there the four models separate cleanly: at low energy it reaches $\simeq11.5$ for Lema\^itre, $\simeq26$ for Hale and Doolen, $\simeq39$ for Viviani and $\simeq46$ for Deltuva --- a factor of four between the extremes. \cref{Spacing}, not \cref{Polarized_DHe3_Reactivity}, is where the QSF uncertainty should be read.

In most cases, the D--D reactivity in a polarized non-thermal plasma lies below that of an unpolarized non-thermal plasma. Since polarization simultaneously raises the deuterium--helium-3 reactivity, the separation between the two should be larger than in the unpolarized case. This is what we find for every QSF evaluation except Lemaître's, whose QSF exceeds unity over the whole energy range considered and therefore raises the D--D reactivity as well.

\par When the reactivity curve for the D--D reaction in the polarized non-thermal regime lies below that of the thermal regime, it leads to an important conclusion: fewer neutrons are produced by the fuel. Strictly speaking, this corresponds to a neutron-lean fuel rather than a strictly aneutronic one, as the D--D branches are suppressed rather than eliminated. 
\par To quantify this difference, we define the Polarized Spacing function:
\begin{equation}
    \mathcal{S}_{pola}(\mathcal{E}_{tot},F)=\frac{\langle \sigma v \rangle_{pola,\ce{D-^3He}}(\mathcal{E}_{tot},F)}{\langle \sigma v \rangle_{pola,D-D}(\mathcal{E}_{tot},F)}
\end{equation}

But we’re also going to define a new function that will generally tell us how much we gain in deviation by polarizing our fuel compared to if we hadn’t polarized it. This function will be the difference between the spread we have in our polarized non-thermal plasma and that in an unpolarized non-thermal plasma, such that if this function is less than 0, it is not worthwhile to polarize in order to differentiate between these two reactivities.

\begin{equation}
    \Delta \mathcal{S}(\mathcal{E}_{tot},F)=\mathcal{S}_{pola}(\mathcal{E}_{tot},F)- \mathcal{S}(\mathcal{E}_{tot},F)
\end{equation}

\par Notice that, for most QSF evaluations in \cref{Spacing}, the spacing is greater when we polarize, except for Lemaître. The effect can be large: in Deltuva’s evaluation, at a hot-tail energy fraction F=0.4, $\mathcal{S}_{pola}$ exceeds 30 below $\mathcal{E}_{tot} \simeq 8 \ \text{keV}$, falling to about 20 at its minimum near 35 keV. 

\par In this section we have seen that, full polarization of the non-thermal reaction not only increases the reactivity of the deuterium helium-3 reaction by a factor of approximately 1.5 compared to the non-thermal reactivity but also tends to increase the gap in term of reactivity between the primary reaction of the $\ce{D-^3He}$ fuel and its D--D secondary reactions in most of the QSF theorized presented. Polarization may also reduce the reactivity of the D--D reactions that produce neutrons, which aligns with the concept of developing a neutron lean reactor.

\section{Ignition Energy} \label{sec:Ignition energy}

\begin{table*}
    \center
    \captionsetup{type=table}
    \caption{Computed ideal ignition energy $E_b$ for polarized and unpolarized non-thermal D-T and $\ce{D^3He}$ fuel, as a function of the hot-tail energy fraction $F$}
    \label{table: polarized Thermal Temperature of ignition}
    
    \begin{tabular}{lcccc}
        \toprule
        \text{F} & \text{\makecell{DT polarized \\  non-thermal \\ $\mathcal{E}_b$ [keV]}} & \text{\makecell{DT unpolarized \\ non-thermal \\  $\mathcal{E}_b$ [keV]}} & \text{\makecell{\ce{D^3He} polarized \\  non-thermal \\ $\mathcal{E}_b$ [keV]}} & \text{\makecell{\ce{D^3He} unpolarized \\ non-thermal \\  $\mathcal{E}_b$ [keV]}}  \\ 
        \midrule
        0.0 & 3.8 & 4.3 & 26.3 & 30.7 \\
        0.1 & $\leq$ 0.5 & 1.0 & 20.0 & 26.7 \\
        0.2 & $\leq$ 0.5 & $\leq$ 0.5 & 9.0 & 19.0 \\
        0.3 & $\leq$ 0.5 & $\leq$ 0.5 & 3.8 & 9.5 \\
        0.4 & $\leq$ 0.5 & $\leq$ 0.5 & 2.1 & 5.1 \\
        0.5 & $\leq$ 0.5 & $\leq$ 0.5 & 1.3 & 3.2 \\
        0.6 & $\leq$ 0.5 & $\leq$ 0.5 & $\leq$ 0.5 & 2.2 \\
        \bottomrule
    \end{tabular}
\end{table*}

\par The ideal ignition energy is defined as the minimum amount of energy that must be supplied to the system to overcome the losses. The only energy losses considered here are the bremsstrahlung radiated when electrons are deflected by ions, which is taken to dominate the other loss channels (\cite{Physique_Des_Tokamaks} shows that other losses are negligible ). 
Just to clarify that what will be done in this section would be a very simple model of Bremsstrahlung losses so it is not an accurate calculation but rather an approach aimed at getting a rough idea of the ideal ignition energy (in term of thermal energy). A more accurate calculation have been done by Xie et al. \cite{Xie_losses}. 
\par First, Bremsstrahlung losses result from the acceleration (or deceleration) of a charged particle (electrons or ions). In plasma, this acceleration is caused by Coulomb collisions between charged particles (electron, deuterium ions, tritium ion, helium-3 ions). In the frame of our study,it is assumed that:
\begin{enumerate}
    \item There is no Coulomb collisions between ions
    \item There is no Coulomb collisions betweens electrons
    \item Ions are not affected by collisions (that means it is no subject to acceleration) from electrons \label{point 3}
    \item Only collisions betweens electrons by ions are considered
    \item There is no relativistic corrections 
\end{enumerate}

\par Point.\labelcref{point 3} can be explained by the fact that the mass of the ion is much more important than the mass of the electron ($\frac{m_{n}}{m_{e}} \approx A \times 2 \times 10^{3} \gg 1$) so due to the significant mass difference, only electron would be affected by the coulomb collision and that the ions are at rest in the electron's referential. 

\par The power radiated (Bremsstrahlung looses) from electrons is given by (derivation in Appendix ~\labelcref{sec: DT total bremsstrahlung power radiation}):
\begin{equation}
    P_{b}(T_e) = \frac{n_i^2 Z^3 e^6}{24\pi^2 \varepsilon_0^3 m_e^{3/2} c^3 \hbar} \sqrt{T_e} \label{eq:final Pb} 
\end{equation}

\par However, this last equation is valid only when two ions Have the same atomic number $Z$ like D--T reaction. For $\ce{D-^3He}$, the total power lost per unit volume is given by (derivation in Appendix~\labelcref{sec: DHe3 total bremsstrahlung power radiation}):

\begin{equation}
    P_{b}(T_{e})=\frac{n_{i}^{2}e^{6}}{96\pi^{2} \varepsilon_{0}^{3}m_{e}^{3/2}c^{3}\hbar}(Z_{D}^{2}+Z_{He}^{2})(Z_{D}+Z_{He})\sqrt{ T_{e} }
\end{equation}

\par Here $T_e$ [keV] is the electron temperature; since electrons follow a Maxwellian velocity distribution, $v_{e^-}$ is given by \cref{v0}.

\par Once the power lost in the plasma have been determined \cref{eq:final Pb} , is now required to determine the power produced from fusion reaction to heat the plasma in order to do a power balance. The power output from fusion stoichiometric reaction is defined by:
\begin{equation}
    P_{fus}(\mathcal{E}_{tot},F)= \frac{n_i^2}{4} \langle \sigma v  \rangle (\mathcal{E}_{tot},F) Q
\end{equation}

For $D-T$ reaction, plasma can be only heated by $\alpha$ particles (which represents 1/5 of energy produced from the reaction), since the neutron is electrically neutral, it does not contribute to heating \cref{DT}, while for $D-^3He$ reaction, both particles created from reaction are charged \cref{DHe3}, therefore power used to heat plasma is:
\begin{align}
    & P_{heating,DT}(\mathcal{E}_{tot},F)=\frac{n_i^2}{20} \langle \sigma v \rangle_{DT}(\mathcal{E}_{tot},F) Q_{DT} \\
    & P_{heating,\ce{D^3He}}(\mathcal{E}_{tot},F)= \frac{n_i^2}{4} \langle \sigma v \rangle_{\ce{D^3He}}(\mathcal{E}_{tot},F) Q_{\ce{D^3He}}
\end{align}

\par Given the large cross section of these fast ions in the hot tail, we can assume that these ions react with a characteristic time shorter than that of collisions between a fast ion and an electron So we assume that our non-thermal distribution is not affected by the presence of electrons. This assumption may be true for small values of F, but may become wrong as F increases.

\begin{center}
    \captionsetup{type=table}
    \caption{Maxwellian-computed ideal ignition temperature compared to the value reported in \cite[Table.3.2, p.36]{Book_Bremsshalung}}
    \label{table: Thermal Temperature of ignition}
    
    \begin{tabular}{lccc}
        \toprule
        \textbf{Reaction} & \textbf{\makecell{reference \\ $T_b$ [keV]}} & \textbf{\makecell{computed \\ $T_b$ [keV]}} & \textbf{deviation} \\ 
        \midrule
        $\ce{D-^3He}$ & 28  & 30.7  & +9.6 \% \\
        $D-T$ & 4.3  & 4.3  & 0 \% \\
        \bottomrule
    \end{tabular}
\end{center}

\begin{center}
    \captionsetup{type=figure, justification=raggedright}
    \centering
    \includegraphics[width=1\linewidth]{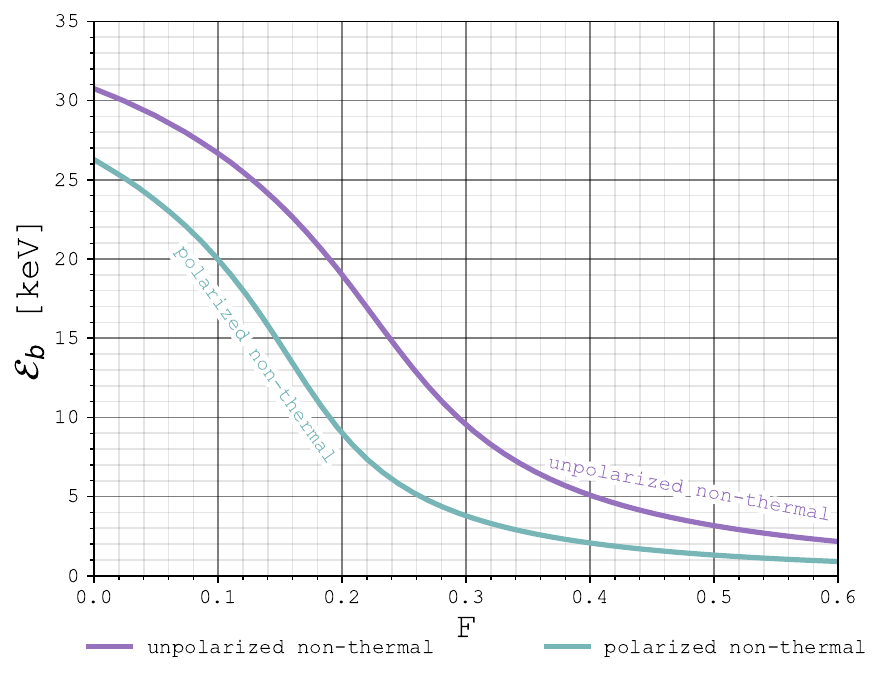}
    \caption{Plot showing the evolution of the ideal ignition energy $\mathcal{E}_b$ for the non-thermal $\ce{D-^3He}$ reaction as a function of the energy fraction in the hot tail $F$, with the teal curve representing the case where the reaction is polarized and the purple curve representing the case where it is non-polarized}
    \label{DHe3Eb}
\end{center}

\par Assuming $T_e=\mathcal{E}_{ions}=\mathcal{E}_{tot}$ (equal distribution of energy between ions and electrons)--- the pessimistic limit for this calculation, since bremsstrahlung grows as $\sqrt{T_e}$ and any real electron population, heated by the fast ions only over a slowing-down time, sits below $\mathcal{E}_{tot}$; the ideal ignition energies below are therefore upper bounds, the ideal ignition energy is denoted $\mathcal{E}_b$ defined by $P_b(\mathcal{E}_b)=P_{heating}(\mathcal{E}_b,F)$. The reference value of ion density in a Tokamak device is $n_i=10^{20} \ m^{-3}$ \cite{Physique_Des_Tokamaks} (This value is provided for context but is not included in the calculation because both powers are proportional in $n_i^2$). The ideal ignition energy are computed in \cref{table: Thermal Temperature of ignition}

\par The expression for bremsstrahlung losses (\cref{eq:final Pb}), evaluated in the thermal case, yields an ideal ignition temperature close to that reported by Xie \cite{Book_Bremsshalung} for D--T, and deviates by 9.6\% for $\ce{D-^3He}$ (see \cref{table: Thermal Temperature of ignition}, where Nevins' formula \cite{Nevins} was used). The complete table of ideal ignition energies can be found in \cref{table: polarized Thermal Temperature of ignition}.

\par Values below the set threshold of $0.5~\text{keV}$ will not be considered, as they represent the lower bound of the Bosch and Hale cross-section parameterization. The non-thermal reaction consists of a thermal component (Maxwellian) and a non-thermal component (hot tail); the limiting component at low energy is the Maxwellian distribution: the temperature does not enter the cross section directly, but it fixes the root-mean-square velocity of the bulk ions and hence the relative velocities at which collisions occur. Therefore, the lower temperature limit in the Maxwellian bulk has been set at $0.5~\text{keV}$, and within this framework, above $F = 0.1$ for polarized non-thermal D--T or $F = 0.2$ for unpolarized non-thermal D--T, there is no minimum ideal ignition energy above the $0.5~\text{keV}$ floor for the D--T reaction. If an extension of the cross-section formula below $0.5~\text{keV}$ is determined, this ideal ignition energy should be very low.
\par For the $\ce{D-^3He}$ reaction, a curve displaying the minimum ideal ignition energy as a function of $F$ has been plotted in \cref{DHe3Eb} . Clearly, the ideal ignition energy is a decreasing function of $F$, with the slope  steeper between $F$=0.15 and $F$=0.3 than beyond that value.

\begin{figure*}[htbp]
    \centering

    \includegraphics[width=1\linewidth]{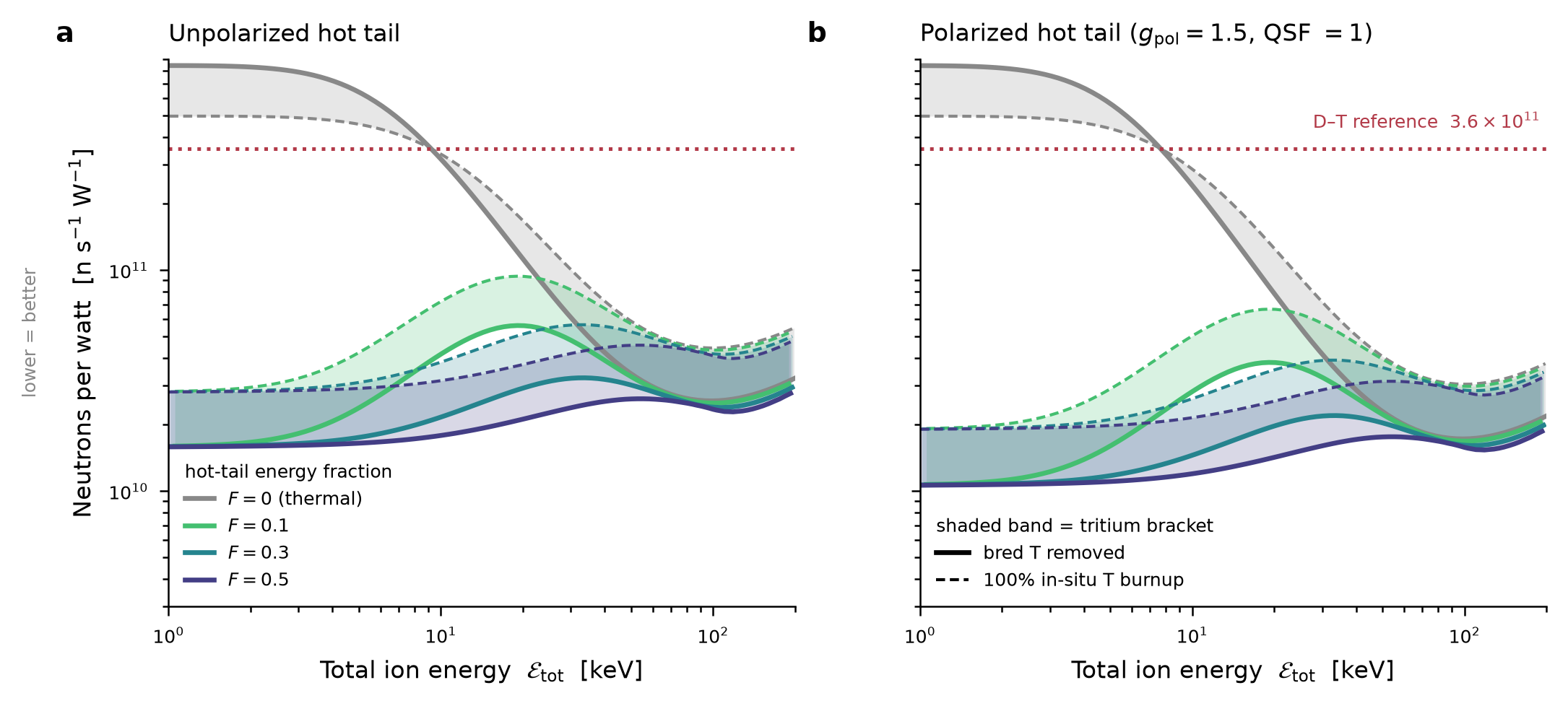}
    \captionsetup{type=figure}
    \caption{Neutrons per watt of total fusion power for a stoichiometric
    \ce{D-^3He} fuel, (a) unpolarized and (b) fully polarized with $g_{pola}=1.5$ and
    D--D QSF $=1$. Solid lines assume bred tritium is removed before burning
    \cref{nW_low}; dashed lines assume $100\%$ in-situ burnup \cref{nW_up}; the shaded
    band is the bracket between them. The dotted line is the D--T value.}
    \label{nWfig}

    \includegraphics[width=1\linewidth]{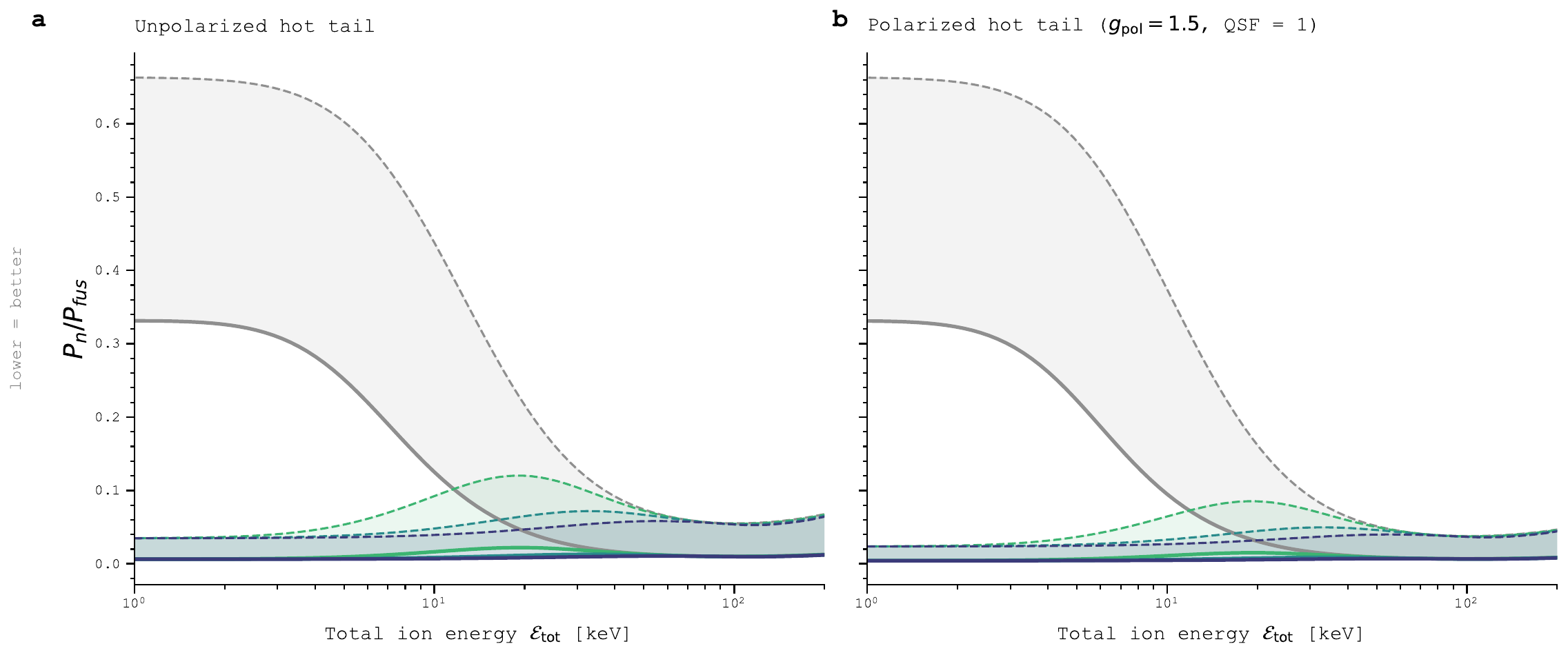}
    \captionsetup{type=figure}
    \caption{$P_n/P_{fus}$ for a stoichiometric
    \ce{D-^3He} fuel, (a) unpolarized and (b) fully polarized with $g_{pola}=1.5$ and
    D--D QSF $=1$. Solid lines assume bred tritium is removed before burning
    \cref{eq:Pn_Pfus_low}; dashed lines assume $100\%$ in-situ burnup \cref{eq:Pn_Pfus_up}; the shaded
    band is the bracket between them. The dotted line is the D--T value.}
    \label{fig:Pn_Pfus}
    
\end{figure*}

This section has shown that the model reduces , by a large factor, the energy required to ignite the $\ce{D-^3He}$ reaction, a reduction that is greater when the fuel is polarized, thereby addressing the problem of high ideal ignition temperatures in the thermal case. This reduction in the energy required to ignite the fuel is also evident for the deuterium-tritium reaction, but the result is less promising because, although it would be easier to initiate the reaction, the operating energies of a reactor using this fuel would still be between 10 keV and 20 keV (see \cref{NT Reactivity}), and we have seen that the hot tail has no impact on the reactivity at those energies.

\section{Neutron-lean performance: neutrons per watt} \label{sec:neutrons per watt}

\begin{figure*}[t]
    \centering
    \captionof{table}{Neutrons per watt at the ideal ignition energy $\mathcal{E}_b$ of
    \cref{table: polarized Thermal Temperature of ignition}. Ranges span the two tritium
    bounds \cref{nW_low,nW_up}. Polarized columns use $g_{\text{pol}}=1.5$ and D--D QSF $=1$.
    The polarized $F=0.6$ entry is omitted: its $\mathcal{E}_b$ falls below the $0.5~\text{keV}$ Maxwellian floor adopted in \cref{sec:Ignition energy}.}
    \label{nWtable}
    \vspace{0.5ex}

    \begin{tabular}{lcccccc}
        \toprule
        & \multicolumn{3}{c}{\textbf{unpolarized}} & \multicolumn{3}{c}{\textbf{polarized}} \\
        \cmidrule(lr){2-4}\cmidrule(lr){5-7}
        \textbf{F} & $\mathcal{E}_b$ [keV] & \makecell{$N/W$ \\ $[10^{10}~\mathrm{n\,s^{-1}\,W^{-1}}]$} & \% of D--T
                   & $\mathcal{E}_b$ [keV] & \makecell{$N/W$ \\ $[10^{10}~\mathrm{n\,s^{-1}\,W^{-1}}]$} & \% of D--T \\
        \midrule
        0.0 & 30.7 & 5.92--9.79 & 16.7--27.6 & 26.3 & 5.06--8.57 & 14.3--24.1 \\
        0.1 & 26.7 & 5.15--8.67 & 14.5--24.4 & 20.0 & 3.82--6.65 & 10.8--18.7 \\
        0.2 & 19.0 & 3.73--6.47 & 10.5--18.2 & 9.0  & 1.69--3.05 & 4.8--8.6   \\
        0.3 & 9.5  & 2.12--3.76 & 6.0--10.6  & 3.8  & 1.13--2.02 & 3.2--5.7   \\
        0.4 & 5.1  & 1.69--2.99 & 4.7--8.4   & 2.1  & 1.08--1.93 & 3.0--5.4   \\
        0.5 & 3.2  & 1.61--2.86 & 4.5--8.0   & 1.3  & 1.06--1.91 & 3.0--5.4   \\
        0.6 & 2.2  & 1.59--2.82 & 4.5--8.0   & --   & --         & --         \\
        \bottomrule
    \end{tabular}

    \vspace{1.5em} 

    \includegraphics[width=\linewidth]{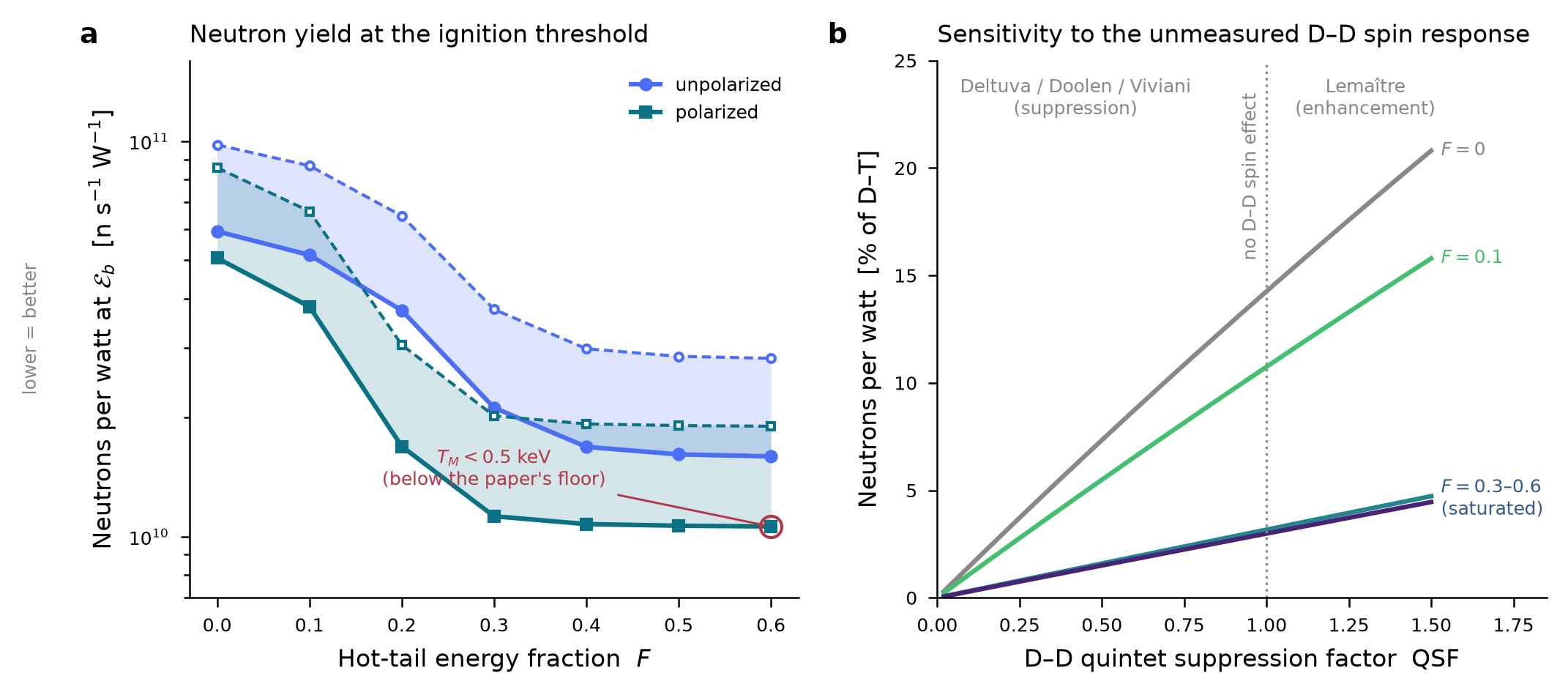}
    \captionsetup{type=figure}
    \caption{(a) Neutrons per watt evaluated at the ideal ignition energy
    $\mathcal{E}_b(F)$ of \cref{table: polarized Thermal Temperature of ignition}; bands
    span the two tritium bounds. (b) Dependence of the polarized metric on the D--D
    quintet suppression factor, at the polarized $\mathcal{E}_b(F)$ where tritons are exhausted.}
    \label{nWfig2}
\end{figure*}

\par The spacing function $\mathcal{S}$ of \cref{spacing} and \cref{Spacing}
establishes that the hot tail separates the primary $\ce{D-^3He}$ reaction from its
D--D secondaries, but a reactivity ratio is not by itself an engineering figure of
merit: what a reactor designer needs is the neutron production per unit of fusion
power. We therefore close the calculation here.

\par \textbf{Neutron inventory.} In a stoichiometric fuel ($n_D=n_{\ce{^3He}}=n_i/2$)
the primary reaction \cref{DHe3} and the D--D proton branch \cref{ddp} are aneutronic;
the only direct neutron source is the D--D neutron branch \cref{ddn}, at $2.45~\text{MeV}$.
The proton branch, however, breeds tritium, which can burn on deuterium and release a
$14.06~\text{MeV}$ neutron \cref{DT}. Because that tertiary channel carries almost six
times the neutron energy of the direct branch, its treatment---not the direct branch---sets
the energy-weighted neutron load, and it depends on a tritium-exhaust model that lies
outside the scope of this work. We therefore bracket it. Writing
$z=\langle\sigma v\rangle_{\ce{D-^3He}}$, $x=\langle\sigma v\rangle_{DD_n}$ and
$y=\langle\sigma v\rangle_{DD_p}$, all evaluated at the same $(\mathcal{E}_{tot},F)$ and
with the shared tail velocity $v_0=v_{opt}(T_M,\ce{D-^3He})$, the volumetric rates are
$2z$, $x$ and $y$ in units of $n_i^2/8$. The two bounds are then:

\begin{align}
&\left.\frac{N}{W}\right|_{\text{T removed}}=\frac{x}{2zQ_{\ce{D^3He}}+xQ_{DD_n}+yQ_{DD_p}}\label{nW_low} \\
&\left.\frac{N}{W}\right|_{\text{full T burn}}=\frac{x+y}{2zQ_{\ce{D^3He}}+xQ_{DD_n}+y(Q_{DD_p}+Q_{DT})} \label{nW_up}
\end{align}

\begin{figure*}[t]
    \centering

    \captionof{table}[Worked example anchored on ARTEMIS: first-wall loading]{%
    Worked example anchored on ARTEMIS \cite{Artemis_DC}: first-wall loading
    for a $200~\text{MW}$ total-fusion-power $\ce{D-^3He}$ chamber with
    $A_w=110~\text{m}^2$, sized by ARTEMIS's own $1.7~\text{MW\,m}^{-2}$ surface-heat limit.
    Ranges span the two tritium bounds \cref{nW_low,nW_up}. Damage is a 14-MeV-equivalent
    upper bound and the last column the time to reach $20~\text{dpa}$. Operating points are
    each case's own ideal ignition energy from
    \cref{table: polarized Thermal Temperature of ignition}. $^{*}$the D--T row is the same
    chamber run on D--T at equal fusion power --- a reference thought experiment, not a D--T
    design point.}
    \label{plantTable}
    \vspace{0.8ex}

    \begin{tabular}{lccccc}
        \toprule
        \textbf{case} & \makecell{$\mathcal{E}_b$ \\ $[\mathrm{keV}]$}
                      & \makecell{$P_n$ \\ $[\mathrm{MW}]$}
                      & \makecell{$q_n$ \\ $[\mathrm{MW\,m^{-2}}]$}
                      & \makecell{damage \\ $[\mathrm{dpa\,fpy^{-1}}]$}
                      & \makecell{to 20 dpa \\ $[\mathrm{fpy}]$} \\
        \midrule
        D--T, same chamber$^{*}$        & 4.3  & 160.0      & 1.454         & 13.57       &  1.5  \\
        thermal $\ce{D-^3He}$ ($F=0$)   & 30.7 & 4.6--25.0  & 0.042--0.227  & 2.26--3.74  & 5   \\
        non-thermal, $F=0.3$            & 9.5  & 1.7--9.4   & 0.015--0.086  & 0.81--1.44  & 14 \\
        non-thermal, $F=0.6$            & 2.2  & 1.3--7.0   & 0.011--0.064  & 0.61--1.08  &  18.5 \\
        polarized non-thermal, $F=0.5$  & 1.3  & 0.8--4.7   & 0.008--0.043  & 0.41--0.73  & 27.5 \\
        \bottomrule
    \end{tabular}

    \vspace{1.5em} 

    \includegraphics[width=0.86\linewidth]{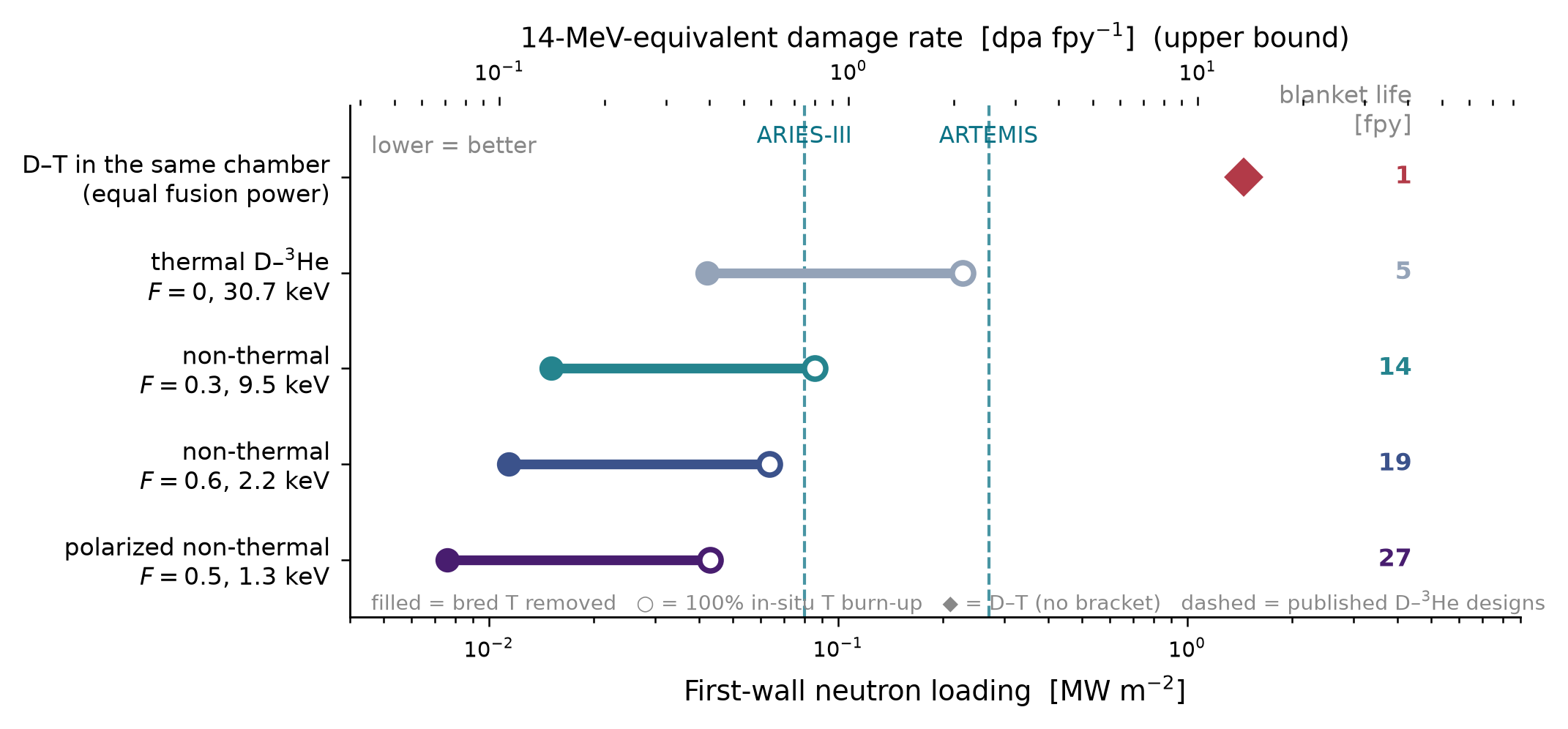}
    \captionsetup{type=figure}
    \caption{First-wall neutron loading for the $200~\text{MW}$ worked example of
    \cref{plantTable}. Filled markers assume bred tritium is removed before burning, open
    markers $100\%$ in-situ burn-up; the bar is the bracket between them. Dashed lines are the
    published loadings of the ARTEMIS \cite{Artemis_DC} and ARIES-III \cite{ARIES3}
    $\ce{D-^3He}$ designs. The upper axis is the 14-MeV-equivalent damage rate and the
    right-hand column the time to reach $20~\text{dpa}$. D--T is a single value because its
    neutron is the primary product.}
    \label{plantFig}
\end{figure*}

\par The lower bound assumes bred tritons are exhausted before they burn; the upper
bound assumes every triton burns in situ, so that in steady state the D--T rate equals the
D--D proton-branch rate and no D--T reactivity need be evaluated. The $\ce{^3He}$ bred by
\cref{ddn} joins the fuel inventory and opens no new neutron channel. Both expressions are
independent of $n_i$, so the metric is a property of the distribution alone. It is
convenient to also quote the neutron \emph{power} fraction
$f_n=P_n/P_{fus}$, obtained by weighting each channel with its neutron energy.

\begin{align}
    &f_n |_{\text{T removed}}= \frac{x\times 2.45 \text{ MeV}}{2zQ_{\ce{D^3He}}+xQ_{DD_n}+yQ_{DD_p}} \label{eq:Pn_Pfus_low} \\
    &f_n |_{\text{full T burn}}= \frac{x \times 2.45 \text{ MeV} +y \times 14.67 \text{ MeV}}{2zQ_{\ce{D^3He}}+xQ_{DD_n}+y(Q_{DD_p}+Q_{DT})} \label{eq:Pn_Pfus_up}
\end{align}

\par For reference, a D--T plasma gives $N/W=3.55\times10^{11}~\mathrm{n\,s^{-1}\,W^{-1}}$ and $f_n=0.80$.

\par \textbf{Results.} \cref{nWfig,fig:Pn_Pfus} plots $N/W$ and $f_n$ over
$\mathcal{E}_{tot}\in[1,200]~\text{keV}$. The thermal case is a useful warning: at
$10~\text{keV}$ an unpolarized Maxwellian $\ce{D-^3He}$ plasma emits $3.2$ -$3.4\times10^{11}~\mathrm{n\,s^{-1}\,W^{-1}}$, i.e. $90$--$95\%$ of the D--T value and $f_n \sim 12\%-42\%$ this explains why, in a thermal $\ce{D-^3He}$ plasma, slightly less than half of the fusion energy is still carried by the neutrons produced by the tertiary D--T reaction, because at that energy the D--D branches outrun the primary reaction
(\cref{Thermal}.b). ``Advanced fuel'' is therefore not synonymous with ``neutron-lean''
at D--T-like operating points. Introducing a hot tail of only $F=0.1$ drops the metric to
$4.2$--$7.2\times10^{10}$, a reduction of $7.6 \times$  (T removed) or $4.7\times$ (full burnup and  a neutron power fraction $f_n \sim 1.6 \%-9.2 \%$; at $F=0.5$ the metric reaches $1.8$--$3.2\times10^{10}$, i.e. $5$--$9\%$ of D--T and $f_n \sim 0.7 \%-3.9 \%$. The curves flatten above $F\simeq0.3$, and the residual $F$-dependence is weakest near
$80~\text{keV}$, where the spread across $F\in[0,0.5]$ narrows to $5\%$; it widens again
to $9\%$ at $100~\text{keV}$ and $16\%$ at $150~\text{keV}$. This tracks the behaviour of
$\mathcal{S}$ (\cref{sec:Non-thermal fusion reaction}), whose own spread is likewise
minimal near $80~\text{keV}$ rather than monotonically vanishing.

\par \cref{nWtable} evaluates the metric at the ideal ignition energies
$\mathcal{E}_b$ of \cref{table: polarized Thermal Temperature of ignition}, which is the
reactor-relevant operating point. The thermal $\ce{D-^3He}$ ignition point
($30.7~\text{keV}$) already gives $17$--$28\%$ of the D--T neutron rate; the polarized
non-thermal case at  $F=0.5$ gives $3.0$--$5.4\%$, a further $5.2$--$5.6\times$ improvement
and $19$--$34\times$ fewer neutrons per watt than D--T. In neutron \emph{power} terms the
same operating point puts only $0.4$--$2.4\%$ of the fusion power into neutrons, against
$80\%$ for D--T. The two effects are not independent: most of the gain comes from the hot
tail, with polarization contributing the remaining factor of $\simeq1.5$ through the
primary reaction.

\par \textbf{What this means for a plant.} The metric above is a property of the
distribution; to make it concrete we apply it to a worked example anchored on a published
$\ce{D-^3He}$ system design rather than on a D--T device. ARTEMIS \cite{Artemis_DC} is a
conceptual $\ce{D-^3He}$ field-reversed-configuration power plant --- the same topology in
which the non-thermal tail of \cref{sec:Non-thermal fusion reaction} has actually been
observed \cite{TAE_non_thermal} --- and reports $1610~\text{MW}$ of fusion power of which
$77~\text{MW}$ is carried by neutrons, a first-wall neutron load of
$0.27~\text{MW\,m}^{-2}$ and a first-wall surface heat load of $1.7~\text{MW\,m}^{-2}$. The
ARIES-III $\ce{D-^3He}$ tokamak study \cite{ARIES3} reaches a comparable neutron fraction of
about $4\%$. Both therefore sit at $f_n\simeq4$--$5\%$, inside the $2.3$--$12.5\%$  bracket
our \cref{nWtable} gives for thermal $\ce{D-^3He}$ at its ignition point --- an independent
check that the metric is calibrated.

\par \textbf{Sizing the chamber.} Because a $\ce{D-^3He}$ plasma releases $95\%$ or more
of its energy in charged particles and radiation, its first wall is limited by \emph{surface
heat flux}, not by neutrons. We therefore size a hypothetical $200~\text{MW}$ (total fusion
power) chamber by holding the first-wall surface load at ARTEMIS's own design value,
$(1-f_n)P_{fus}/A_w\leq1.7~\text{MW\,m}^{-2}$, which gives $A_w\simeq110~\text{m}^2$ --- an
FRC of wall radius $1.4~\text{m}$ and length $12.5~\text{m}$. All rows of \cref{plantTable}
then share one fixed geometry, so they differ only through the distribution. The neutron
source rate is $S_n=(N/W)P_{fus}$, the neutron power $P_n=f_nP_{fus}$, and the wall
quantities follow as $\Gamma_w=S_n/A_w$ and $q_n=P_n/A_w$; everything scales linearly with
$P_{fus}$ and inversely with $A_w$.

\par For the damage rate we use the standard steel conversion, $1~\text{dpa}$ per
$1.5\times10^{24}~\text{n\,m}^{-2}$ at $E_n=14~\text{MeV}$, applied to the \emph{total}
neutron current. Because the $\ce{D-^3He}$ neutrons are predominantly the $2.45~\text{MeV}$
D--D branch, whose displacement cross section in iron is roughly three times smaller per
neutron than at $14~\text{MeV}$, the figures below are conservative upper bounds \cite{dpa_Fe}. The last
column converts this to the time needed to reach the $20~\text{dpa}$ limit conventionally
adopted for a first replaceable blanket.

\par The spread changes engineering conclusions rather than merely decorating them
(\cref{plantFig}). Run on D--T, this chamber would see $160~\text{MW}$ of neutron power,
$1.45~\text{MW\,m}^{-2}$ and $\simeq14~\text{dpa\,fpy}^{-1}$, reaching the replacement limit
in about $1.5$ full-power years. A \emph{thermal} $\ce{D-^3He}$ fill at its own ignition
point already improves this to $0.042$--$0.23~\text{MW\,m}^{-2}$ and a $5~\text{fpy}$ blanket
life. ARTEMIS reports $0.27~\text{MW\,m}^{-2}$, so our upper bound sits $16\%$ below it rather
than bracketing it --- close agreement for two independent calculations at different operating
points and fuel mixes, but not a containment. But that is where
a thermal advanced fuel stops. Adding the hot tail is what moves the needle: at $F=0.3$ the
loading falls to $0.015$--$0.086~\text{MW\,m}^{-2}$ ($14~\text{fpy}$), and the polarized
non-thermal case at $F=0.5$ reaches $0.008$--$0.043~\text{MW\,m}^{-2}$ with a first wall that
survives $27~\text{fpy}$ --- longer than the plant. Neutron power falls from $160~\text{MW}$
to $0.8$--$4.7~\text{MW}$.

\par Two consequences are worth drawing out. First, first-wall replacement stops being a
scheduled maintenance activity and becomes a once-per-lifetime event, which removes the
remote-maintenance driver that dominates D--T plant design. Second, the engineering problem
inverts: in this chamber D--T carries four times more neutron load than surface heat load,
whereas every $\ce{D-^3He}$ row carries $20$--$200$ times more surface heat
($\simeq1.8~\text{MW\,m}^{-2}$) than neutron load. For a neutron-lean fuel the first wall is
a heat-flux problem and the neutron inventory is a shielding afterthought --- the opposite of
the D--T case. Note finally that these are equal-\emph{fusion}-power comparisons; with $95\%$
of the output in charged particles, the same $200~\text{MW}$ yields considerably more
electrical power through direct conversion \cite{Artemis_DC}.

\par \textbf{Sensitivity to the D--D spin response.} Because the D--D quintet
suppression factor enters both numerator and denominator of \cref{nW_low,nW_up} almost
linearly, $N/W$ is close to proportional to QSF (\cref{nWfig2}.b). We therefore report it
as a swept parameter rather than adopt one of the four evaluations of \cref{all QSF}: at
$F=0.3$ the polarized metric runs from $0.8\%$ of D--T at QSF $=0.25$ to $4.7\%$ at
QSF $=1.5$, so the spread between the Deltuva-type suppression and the Lemaître-type
enhancement is a factor of $\simeq6$ in neutron output. This is the single largest
uncertainty in the present estimate and is precisely what PolFusion
\cite{The_status_of_double_polarized_dd,View_on_Polfusion} is expected to settle.

\par \textbf{Caveats.} Four limitations apply. (i) The fuel is held stoichiometric;
a $\ce{^3He}$-rich mix would suppress the D--D branches further and is not explored here.
(ii) T--T, $\ce{^3He}$--$\ce{^3He}$ and $\ce{^3He}(n,p)$T channels are neglected, as are
neutrons from beam--wall interaction. (iii) The bred triton is assumed either removed or
fully burnt; partial burnup interpolates between the bounds. (iv) The metric is referred
to total fusion power; referring it to directly convertible charged-particle power would
lower the D--T reference and raise the $\ce{D-^3He}$ advantage further.

\section{Conclusion}
\par Our study addressed some of the shortcomings that could be attributed to the advanced $\ce{D-^3He}$ fuel by polarizing our fuel in a non-thermal regime. 
\par The first issue was its low reactivity compared to the D-T reactions and the fact that, over a certain energy range, the reactivity of the D-D reaction could outweigh that of the $\ce{D-^3He}$ reaction (see \cref{Thermal}). Operating the fuel in a non-thermal regime allowed this latter reaction (much more so than the other three considered) to achieve a massive increase in reactivity about one order of magnitude at 10 keV with a total energy proportion in the hot tail $F$ of 10 \%, up to about two orders of magnitude at 60 \%. This enhancement diminishes at 20 keV but remains significant nonetheless, about a factor of 2 at 10 \% to a factor of 10 at 60 \% (the range [10 keV, 20 keV] is the energy range corresponding to the temperatures in a D-T fusion reactor) \cref{Rrate}. Unlike the other reactions, which reach at most a factor of about 3 for a 60 \% fraction at 10 keV (see \cref{Rrate}).

\par There is therefore an asymmetric reactivity enhancement of the non-thermal reaction. The enhancement of the $\ce{D-^3He}$ reaction increases more than that of the D-D reactions at a given energy, visually spreading the reactivity curves from one to another and thereby allowing, for a given output power, a reactor to be neutron lean (see \cref{spacing}). The polarization of this fuel increases the reactivity of the D-He3 reaction by a factor of roughly 1.5 across the entire energy range, whereas the polarization of D-D reactions in the fuel has been studied in light of four theories that predict either a reactivity gain like  DT and D-He3 reactions, or in contrary, a reactivity loss \cite{Doolen,Lemaitre,Deltuva,Viviani}. Depending on the situation, polarization may or may not be beneficial, given the goal of achieving a greater spread in these reactivity which is advantageous in terms of reducing the number of produced neutrons (see \cref{Spacing}).
\par The second issue with the $\ce{D-^3He}$ reaction is its high ideal ignition energy (Temperature of ideal ignition), which is around 30 keV in thermal conditions, but we were able to address this issue because our non-thermal reaction could, in theory, reduce this ideal ignition energy from 30 keV in thermal conditions to 9.5 keV with 30 \% energy fraction in the hot tail and down to 2.2 keV for a 60 \% fraction. The polarization of this fuel further accentuates this decrease, ranging from 26.3 keV for a 0 \% fraction (polarized fuel in thermal regime), through 3.8 keV for a 30 \% fraction, down to 1.3 keV for a 50 \% fraction (see \cref{table: polarized Thermal Temperature of ignition} and \cref{DHe3Eb}).
Thus, our non-thermal regime addresses two points: a reduction in neutron production at a given power and a reduction in the energy required to initiate the reaction, with polarization further accentuating these phenomena. Quantitatively (\cref{sec:neutrons per watt}), the first of these amounts to a $5.2$--$5.6\times$ reduction in neutrons per watt relative to the thermal $\ce{D-^3He}$ ignition point, and $19$--$34\times$ relative to D--T. For a $200~\text{MW}$ plant sized on the ARTEMIS design (\cref{plantTable}) that is a first-wall neutron loading of $0.008$--$0.043~\text{MW\,m}^{-2}$ rather than $1.45~\text{MW\,m}^{-2}$, and a blanket that lasts $27$ full-power years rather than $1.5$. However, it is important to keep in mind that this study is purely theoretical, feasibility is a different matter entirely, because, on the one hand, the non-thermal “hot tail” reaction that has been observed experimentally \cite{TAE_non_thermal} is not as well controlled as the theory would suggest, the lack of data on the polarized DD reaction prevents us from reaching a clear conclusion, and finally, the survival of a polarized plasma is also a significant challenge.

\section*{Discussion}
A recent paper by Parisi et al. \cite{Parisi_DHe3} entitled "Revisiting Fusion in D-3He Plasmas With Spin-Polarized Fuel", bears directly on this work. This paper examines how polarization might affect the DHe3 reaction, particularly in light of the polarization of secondary D-D reactions that lead to tertiary reactions in terms of power output. Here, the focus is solely on the reactivity specific to each reaction, examining how the non-thermal reaction "spread" the reactivity of DHe3 from that of D-D reactions within an energy range, and whether or not polarisation might tend to accentuate this phenomenon.
It is important to keep in mind that the increase in reactivity for the reactions considered may be much greater than in the context outlined, because the $\alpha$-channelling phenomenon described in \cite{parisiEnhancedAlphaChanneling2026} has displayed that polarization can enhance the $\alpha$-channelling effect, allowing ions to be heated to a greater extent than electrons, thereby adding a reactivity gain from thermal effects in addition to the gain from quantum effects.
\par Our calculation of ideal ideal ignition energy is extremely imprecise; it merely provides a rough idea of the trend and the approximate value we should expect. Therefore, there is a need to refine this calculation by eliminating certain assumptions.
\par We have seen that the gap between the curves for the main $\ce{D-^3He}$ reaction and its secondary D--D reactions in the $\ce{D-^3He}$ fuel, created by the non-thermal regime is positive and can be significant, which would lead to reduce neutron production at a fixed power output. \cref{sec:neutrons per watt} now carries out that calculation. Counting the direct D--D neutron branch \cref{ddn} and, as an upper bound, the $14.06~\text{MeV}$ neutron from in-situ burnup of the tritium bred by \cref{ddp}, the neutrons-per-watt metric falls from $17$--$28\%$ of the D--T value at the thermal $\ce{D-^3He}$ ignition point to $3.0$--$5.4\%$ for the polarized non-thermal fuel at $F=0.5$. The corresponding neutron power fraction is $0.4$--$2.4\%$, against $80\%$ for D--T.

\section*{Acknowledgments}
Many thanks to Timothée Nicolas \orcidlink{0000-0003-1343-7941} for his paper work, from which the “hot tail” non-thermal reaction is taken, and for his help in reproducing some of his figures.
This present work was supported in the context of an internship at the Princeton Plasma Physic Laboratory, NJ, USA, while being a student at IMT Atlantique, France.

\section*{Data Availability and Calculation Detail}
Detailed calculations and supporting datasets are available upon request from Pascal Zhang.

\end{multicols}

\newpage
\section*{Appendix}

\appendix

\counterwithin{equation}{section} 
\counterwithin{figure}{section}
\addtocontents{toc}{\protect\setcounter{tocdepth}{0}}

\begin{multicols}{2}

\section{Constant polarized enhancement factor for $D-T$ and $\ce{D^3He}$ reaction, $g_{pola}=1.5$ } \label{sec:constant gain}

\begin{figure*}[t]
    \captionsetup{type=figure}
    \centering
    \includegraphics[width=1\linewidth]{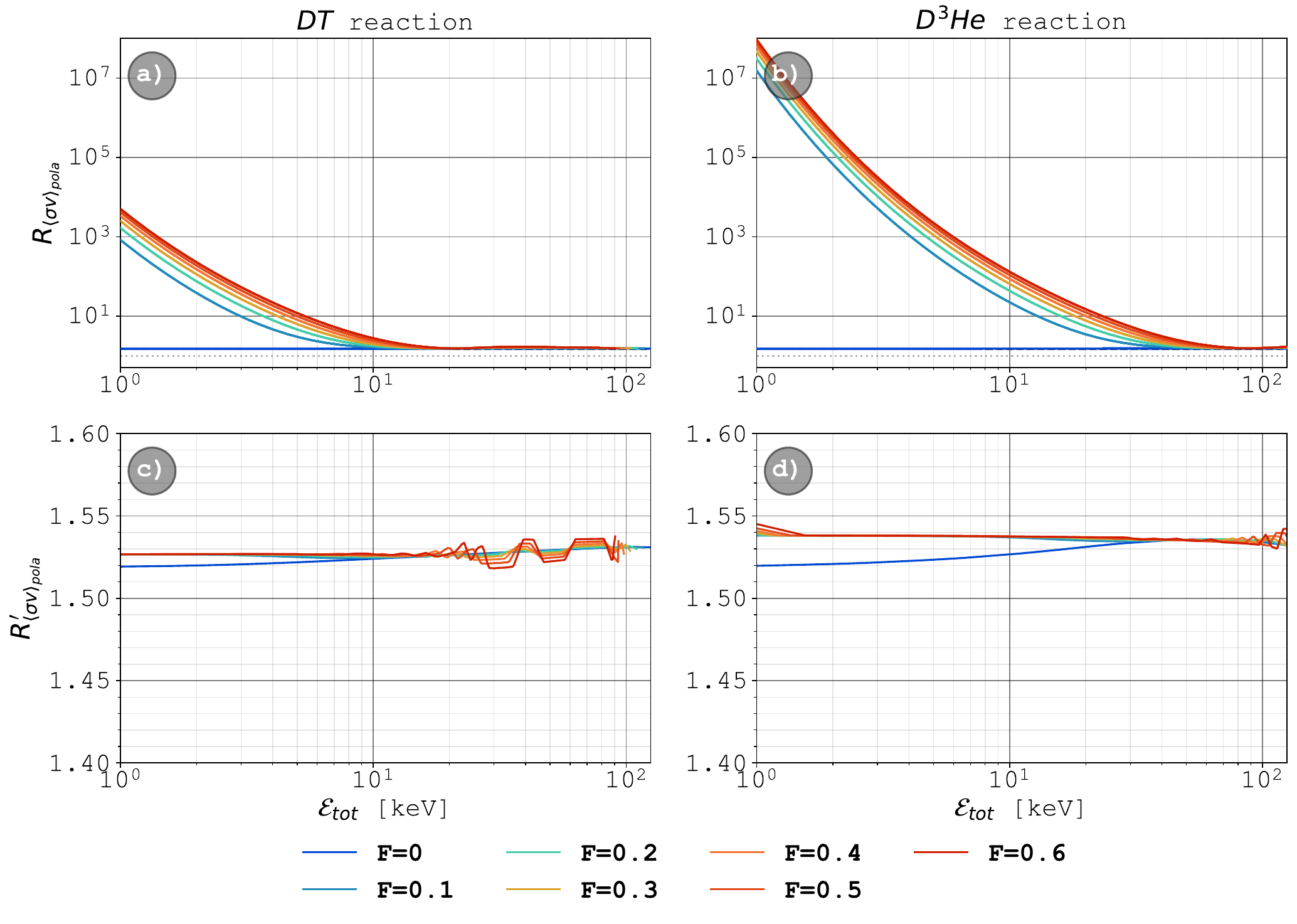}
    \caption{Reactivity increase rate of the polarized non-thermal reaction to thermal $R_{\langle \sigma v \rangle_{pola}}$ (top)(a,b) and polarized non-thermal to unpolarized non-thermal $R^{'}_{\langle \sigma v \rangle_{pola}}$ (bottom)(c,d) for D-T and $\ce{D^3He}$ reactions}
    \label{fig:constant factor}
\end{figure*}

\par Then $\underset{p_z,q_z,p_{zz}}{\max} g_{pola}$ have been plotted in \cite[Fig.4b]{Ab_initio_DT}. This plot has been reproduced and fitted  in the range of $[0 \ keV,1000 \ keV]$ of centre-of-mass energy outside this range, the fit is held constant at its boundary value (see \cref{Delta}).

\begin{center}
    \captionsetup{type=figure}
    \centering
    \includegraphics[width=1\linewidth]{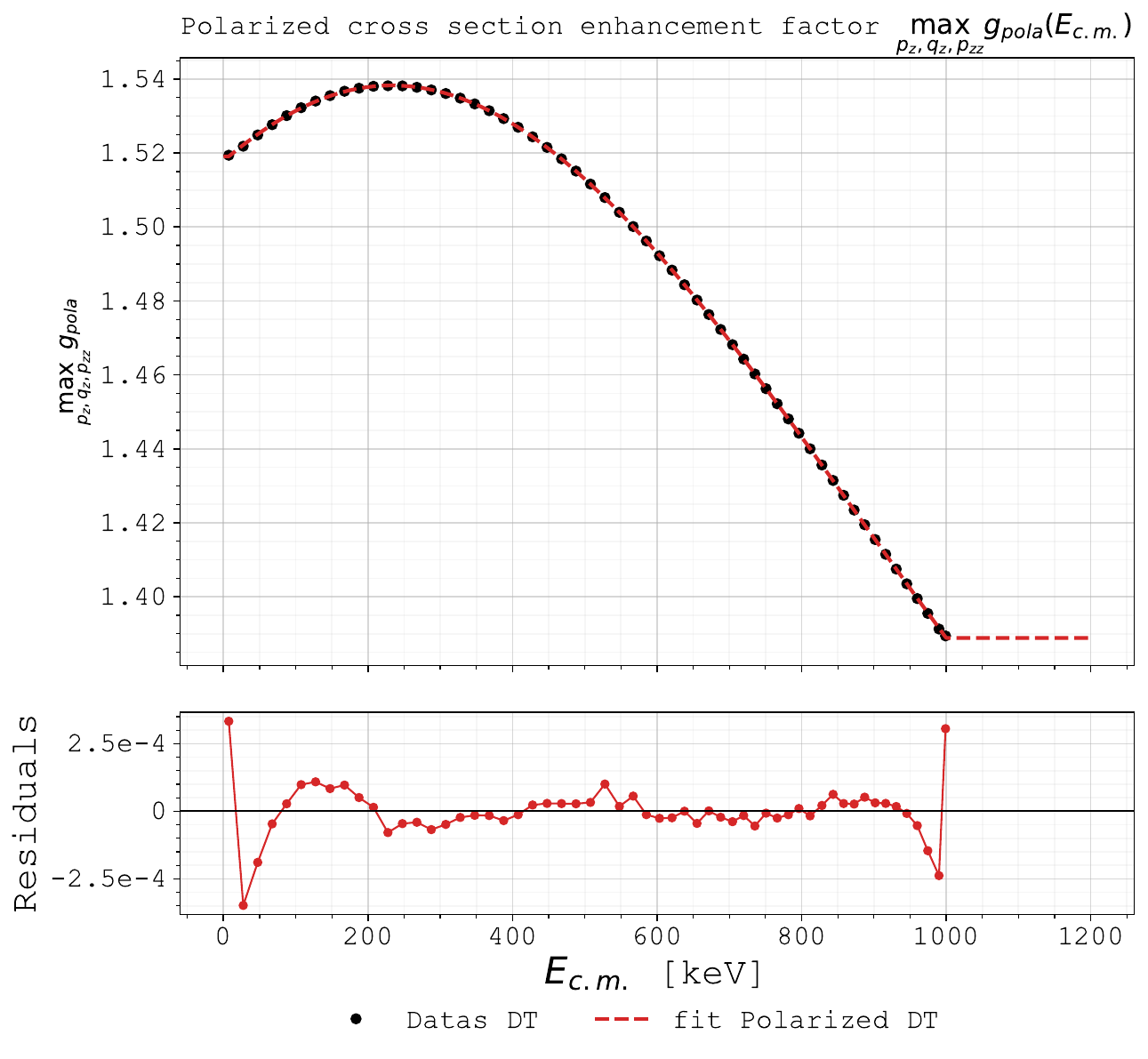}
    \caption{Computed and fitted polarized cross-section enhancement factor for the D--T reaction from Hupin, Quaglioni, and Navr\'atil (2019) \cite[Fig.~4b, p.~5]{Ab_initio_DT}. The curve is fitted between $0~\text{keV}$ and $1000~\text{keV}$ in center-of-mass energy; outside this range, it is held constant at its boundary values}
    \label{Delta}
\end{center}

\par Two caveats apply to this fit. First, the original figure in \cite{Ab_initio_DT} has a non-linear $y$-axis, so the digitisation is imperfect: we obtain $g_{pola}(E_{\text{c.m.}}=0) \approx 1.52$ instead of the exact value $1.50$. Second, the enhancement factor was computed only for the D--T reaction, and we assume that it applies equally to the $\ce{D-^3He}$ reaction. This is justified by the fact that tritium and helium-3 are mirror nuclei, so a similar spin response is expected---though shifted in energy, since the resonant state $J=3/2$ of $\ce{D-^3He}$ is displaced relative to that of D--T \cite{discussion_DT}. The results below should therefore be read as trends rather than as quantitatively accurate values for the D--T and $\ce{D-^3He}$ reactions.

\par Let's revisit the equations we derived earlier for our non-thermal reaction and see how things might change if we induce fully a polarization ($p_zq_z=1$)

Cross section can be express by (see \cref{Polarized Cross Section}):
\begin{equation}
    \sigma_{pola}(E_{c.m})=g_{pola}(E_{c.m}) \sigma_{unpol}(E_{c.m})
\end{equation}

Then reactivity can be expressed as:

\begin{equation}
\begin{split}
    \langle \sigma v \rangle_{pola}(T_M,F)=\int_{0}^{\infty}g_{pola}(u)&\sigma_{unpol}(u)u  \\
    & \times h(u,T_M,F,v_{0,pola}) du
\end{split}
\end{equation}

and with:
\begin{equation}
    v_{opt,pola}(T_{M})=\underset{v_{0}}{arg \ max} \ \frac{\partial \langle\sigma v \rangle_{pola}}{\partial F}_{|\mathcal{E}_{tot}=C^{te},F=0}(T_{M})
\end{equation}

The expression of the optimal energy is different for a non-thermal polarized fuel, this equation is plotted in \cref{Polarized Optimal energy 3N} 

\begin{center}
    \captionsetup{type=figure}
    \centering
    \includegraphics[width=1\linewidth]{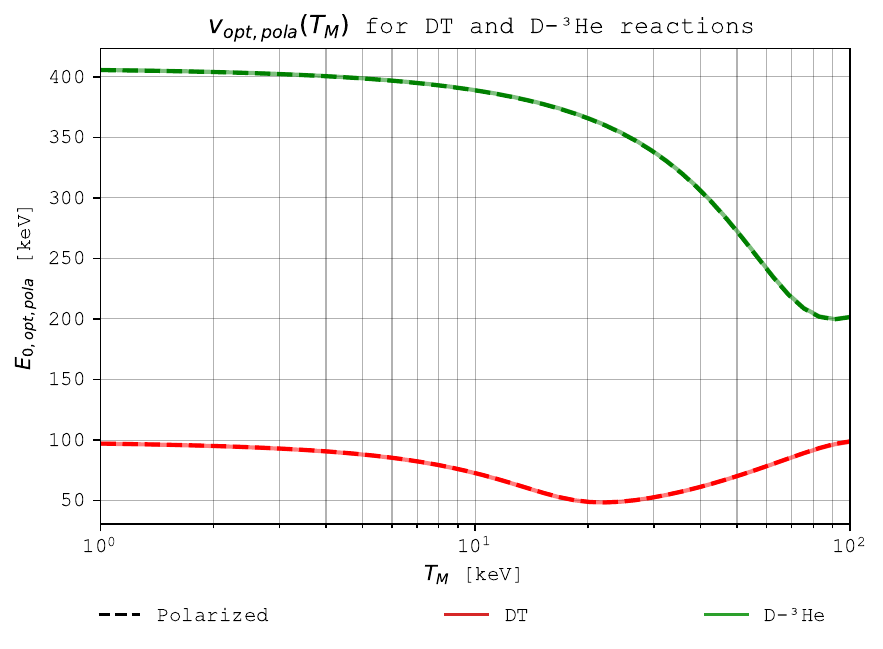}
    \caption{Optimal deuterium energy, $E_{D,\text{opt,pola}}(T_M) = \frac{1}{2}m_D v_{\text{opt,pola}}^2(T_M)$, for DT and D-$^3$He reactions as a function of $T_M$.Dashed lines represent polarized optimal energy, while transparent curves show the corresponding unpolarized optimal energy}
    \label{Polarized Optimal energy 3N}
\end{center}

The polarized optimal velocity overlaps the unpolarized one almost exactly, because $g_{pola}$, whose shape is given in \cref{Delta}, varies very little over the relevant range. Between zero and the cross-section peaks listed in \cref{table Ed}, $[0, 64]~\text{keV}$ for D--T and $[0, 250]~\text{keV}$ for $\ce{D-^3He}$, the slope of $g_{pola}$ is positive but small: the peak of $g_{pola}$ at $\approx 200~\text{keV}$ lies only $0.02$ above its value at $\approx 0~\text{keV}$, whereas the unpolarized cross section varies by several orders of magnitude over the same range.

Let's define two new ratios, the one is the increase rate of the polarized non-thermal reaction to thermal:
\begin{equation}
    R_{\langle \sigma v \rangle_{pola}}=\frac{\langle\sigma v \rangle_{pola}(\mathcal{E}_{tot},F)}{\langle\sigma v \rangle_{M}(\mathcal{E}_{tot})}
\end{equation}

the second ratio, the increase rate of the polarized non-thermal reaction to unpolarized non thermal. 
\begin{equation}
    R^{'}_{\langle \sigma v \rangle_{pola}}=\frac{\langle\sigma v \rangle_{pola}(\mathcal{E}_{tot},F)}{\langle\sigma v \rangle(\mathcal{E}_{tot},F)}
\end{equation}

\par This new increase rate will help us to determine whether yes or no, we could consider a constant increase independent of the energy by polarizing our non-thermal reaction. As we see in \cref{fig:constant factor}.c and \cref{fig:constant factor}.d, the increase rate of the polarized non-thermal reaction to unpolarized non thermal is roughly equal to a constant $R^{'}_{\langle \sigma v \rangle_{pola}} \approx C^{te}$ for both reactions. This constant close to, but not equal to, 1.5: the figure gives $\simeq 1.527$ for D--T and $\simeq1.538$ for $\ce{D-^3He}$. For the D--T reaction, one might think that it doesn't matter if we apply the offset of +0.02 at $E_{c.m} \approx 0$. But here we will simply take the hypothetical maximum attainable value and the constant we want it to equal. If we assume $R^{'}_{\langle \sigma v \rangle_{pola}} = 1.5$ and the maximum achievable value is 1.545, the relative error would be 3 \%. But does that mean we can say that:
\begin{equation}
    g_{pola}=1.5 \Longleftrightarrow R^{'}_{\langle \sigma v \rangle_{pola}} = 1.5  
\end{equation}

\par Let's prove that. First, rewrite the non-thermal reactivity:
\begin{equation}
\begin{split}
    \langle \sigma v \rangle_{pola}(T_M,F)=\int_{0}^{\infty}g_{pola}(\cancel{u})&\sigma_{unpol}(u)u  \\
    & \times h(u,T_M,F,v_{opt,pola}) du
\end{split}
\end{equation}

We have already seen that $v_{opt,pola}=v_{opt}$ in \cref{Polarized Optimal energy 3N} but let's prove that by calculation:
\begin{align}
\langle \sigma v \rangle_{pola}&= \int_{0}^{+\infty} \sigma(u)_{pola}uh(u,T_M,F,v_{0})du \\
&=g_{pola}(\cancel{u})\int_{0}^{+\infty} \sigma(u)_{unpola}uh(u,T_M,F,v_{0})du \\
&=g_{pola} \langle \sigma v \rangle_{unpola}
\end{align}

\begin{align}
v_{\text{opt,pol}}(T_M) &= \operatorname*{arg\,max}_{v_0} \left. \frac{\partial \langle\sigma v \rangle_{pol}}{\partial F} \right|_{\substack{\mathcal{E}_{\text{tot}}=C^{te} \\ F=0}} (T_M) \\
&= \operatorname*{arg\,max}_{v_0} g_{pol} \left. \frac{\partial \langle\sigma v \rangle_{unpol}}{\partial F} \right|_{\substack{\mathcal{E}_{\text{tot}}=C^{te} \\ F=0}} (T_M) \\
&= v_{\text{opt}}(T_M)
\end{align}

Then:
\begin{align}
\langle \sigma v \rangle_{pola}&=g_{pola} \langle \sigma v \rangle_{unpola} \\
&=R^{'}_{\langle \sigma v \rangle_{pola}} \langle \sigma v \rangle_{unpola}
\end{align}

So
\begin{equation}
    \boxed{g_{pola}=1.5 \implies R^{'}_{\langle \sigma v \rangle_{pola}} = 1.5}
\end{equation}

Then, let's begin by $R^{\prime}_{\langle \sigma v \rangle}$:

\begin{align}
	R^{\prime}_{pola}&=\frac{\langle \sigma v \rangle_{pola}(\mathcal{E}_{tot},F)}{\langle \sigma v \rangle (\mathcal{E}_{tot},F)} \\
	 \langle \sigma v \rangle_{pola}(\mathcal{E}_{tot},F)&=R^{\prime}_{pola}\langle \sigma v \rangle (\mathcal{E}_{tot},F) \\
\end{align}

\begin{align}
    &\int \sigma_{pola}(u)uh(u,T_{M},F,v_{opt,pola})du \\
    &=R^{\prime}_{pola}\int \sigma_{unpol}(u)u h(u,T_{M},F,v_{opt})du 
\end{align}

Since $v_{opt,pola}=v_{opt}$ (\cref{Polarized Optimal energy 3N})

\begin{align}
    &\int g_{pola}(u)\sigma_{unpol}(u)uh(u,T_{M},F,v_{opt,\cancel{pola}})du \\
    &=\int R^{\prime}_{pola}\sigma_{unpol}(u)uh(u,T_{M},F,v_{opt})du
\end{align}

So,
\begin{equation}
    \boxed{g_{pola}=1.5 \Longleftarrow R^{'}_{\langle \sigma v \rangle_{pola}} = 1.5}
\end{equation}

Finally,
\begin{equation}
    \boxed{g_{pola}=1.5 \Longleftrightarrow R^{'}_{\langle \sigma v \rangle_{pola}} = 1.5} 
\end{equation}

Without losing much precision, we could consider that $g_{pola}=C^{te} = 1.5$ within our energy range [0,125] keV at the cost of a systematic underestimate of about 2-3 \%

\section{D--T total bremsstrahlung power radiation} \label{sec: DT total bremsstrahlung power radiation}

\par The Larmor formula gives the power radiated by an electron undergoing an acceleration $a$:
\begin{equation}
    P_{L}=\frac{e^{2}a^{2}}{6\pi \varepsilon_{0}c^{3}}
\end{equation}

To evaluate this acceleration undergo by a unique electron, apply the Newton's second law with coulomb collisions.
\begin{equation}
    a^{2}=\frac{e^{2}}{m_{e}^2}\langle | \vec{E}_{tot} |^{2} \rangle
\end{equation}

Let's be focus now about the interaction between a unique electron with a unique ion. At a distance r, a ion of charge $+Ze$, the norm of electric field generated from the ion is given by:
\begin{equation}
    E_{ion}(r)=\frac{Ze}{4\pi\varepsilon_{0}r^{2}}
\end{equation}

Since the Larmor power depends on the square of the acceleration, we need the square of the field:
\begin{equation}
    E_{ion}^2(r) = \frac{Z^2 e^2}{16\pi^2 \varepsilon_0^2 r^4}
\end{equation}

Our electron is at the center. It is immersed in the total electric field $\vec{E}_{tot}$, which is the vector sum of the fields of all the surrounding ions: $\vec{E}_{tot} = \sum \vec{E}_{i}$. The quantity of interest for energy is the spatial average of the square of this field: $\langle \vert{}\vec{E}_{tot}\vert{}^2 \rangle$. When you square a vector sum, you obtain the scalar terms ($\vec{E}_i \cdot \vec{E}_i$) and the cross terms ($2 \vec{E}_i \cdot \vec{E}_j$). Since the ions are distributed purely at random, the cross-terms cancel each other out on average (for every ion pulling in one direction, there is, on average, another pulling in the opposite direction). Then transform this discrete sum into a continuous integral over the entire volume, using  spherical shells, each containing $dN = n_i (4\pi r^2 dr)$ ions ($n_i=n_D+n_{3N}$):

\begin{equation}
    \langle \vert{}\vec{E}_{tot}\vert{}^2 \rangle = \int_{\lambda_c}^{+\infty} E_{ion}^2(r) \times (n_i 4\pi r^2) dr
\end{equation}

where $\lambda_{c}$ is the electron de Broglie wavelength, used as the lower cut-off of the impact-parameter integral. We substitute the square of the field into our integral:
\begin{equation}
    \langle \vert{}\vec{E}_{tot}\vert{}^2 \rangle = \int_{\lambda_c}^{+\infty} \left( \frac{Z^2 e^2}{16\pi^2 \varepsilon_0^2 r^4} \right) n_i 4\pi r^2 dr
\end{equation}

Then:
\begin{equation}
    \langle \vert{}\vec{E}_{tot}\vert{}^2 \rangle = \frac{n_i Z^2 e^2}{4\pi \varepsilon_0^2 \lambda_c}
\end{equation}

acceleration can be now derived into:
\begin{equation}
    a^2 = \frac{n_i Z^2 e^4}{4\pi \varepsilon_0^2 m_e^2 \lambda_c}
\end{equation}

We substitute this acceleration directly into the Larmor formula to find the power radiated by our single electron:
\begin{equation}
    P_{1e^{-}} = \frac{n_i Z^2 e^6}{24\pi^2 \varepsilon_0^3 m_e^2 c^3 \lambda_c}
\end{equation}

To find the total power lost per unit volume, simply multiply the power loss of a single electron by the number of electrons per cubic meter ($n_e$):
\begin{equation}
    P_{b} = \frac{n_i n_e Z^2 e^6}{24\pi^2 \varepsilon_0^3 m_e^2 c^3 \lambda_c} \label{Pb}
\end{equation}

\cref{Pb} which has been found is very similar to one of the equation derived from \cite{Physique_Des_Tokamaks} even if there is a factor $4/\pi$ of difference. Then \cref{Pb} can be developed with 
\begin{align}
    &\lambda_c= \frac{\hbar}{m_ev_{e^-}} \\
    & v_{e^-} \approx \sqrt{\frac{T_e}{m_e}} \quad \textbf{thermal distribution} \label{v0} \\ 
    &n_e= Z n_i \quad \textbf{50:50 fuel}
\end{align}
    
Finally,
\begin{equation}
\boxed{P_{b}(T_e) = \frac{n_i^2 Z^3 e^6}{24\pi^2 \varepsilon_0^3 m_e^{3/2} c^3 \hbar} \sqrt{T_e}} 
\end{equation}

\section{$\ce{D-^3He}$ total bremsstrahlung power radiation} \label{sec: DHe3 total bremsstrahlung power radiation}

Define, $\vec{E}_D$ and $\vec{E}_{He}$ the electric field generated by the deuteron and the helium-3 ion respectively. The total electric field is given by:
\begin{equation}
    \vec{E_{tot}}=\sum_{i} \vec{E}_{D,i}+\sum_{j} \vec{E}_{He,j}
\end{equation}

Consequently, the squared magnitude of the total electric field is given by:
\begin{align}
|\vec{E}_{tot}|^{2}&= \vec{E}_{tot} \cdot \vec{E_{tot}} \\
\begin{split}
    &=\left( \sum_{i} \vec{E}_{D,i} \right)^{2}+\left( \sum_{j} \vec{E}_{He,j} \right)^{2} \\
    &\quad +2\sum_{i} \vec{E}_{D,i}\sum_{j} \vec{E}_{He,j} 
\end{split}
\\
\begin{split}
    &=\sum_{i} \vec{E}_{D,i}^{2}+\sum_{j} \vec{E}_{He,j}^{2}+\sum_{i \neq k} \vec{E}_{D,i} \cdot \vec{E}_{D,k} \\
    & \quad +\sum_{j \neq l} \vec{E}_{He,j} \cdot \vec{E}_{He,l}+2\sum_{i,j} \vec{E}_{D,i} \cdot \vec{E}_{He,j}
\end{split}
\end{align}

Then, the average value $\langle |\vec{E}_{tot} | \rangle$ (Based on the same symmetry criteria as in \cref{sec:Ignition energy}):
\begin{equation}
    \langle \vert{}\vec{E}_{tot}\vert{}^2 \rangle = \left\langle \sum_{i=1}^{N_D} \vert{}\vec{E}_{D,i}\vert{}^2 \right\rangle + \left\langle \sum_{j=1}^{N_{He}} \vert{}\vec{E}_{He,j}\vert{}^2 \right\rangle
\end{equation}

with:
\begin{equation}
	E_{D}^{2} = \frac{Z_{D}^{2} e^{2}}{16 \pi^{2} \varepsilon_{0}^{2} r^{4}} \qquad E_{He}^2(r) = \frac{Z_{He}^2 e^2}{16\pi^2 \varepsilon_0^2 r^4}
\end{equation}

Then transform this discrete sum into a continuous integral overt the entire volume, using spherical shells:
\begin{align}
    \begin{split}
        \langle \vert{}\vec{E}_{tot}\vert{}^2 \rangle &= \int_{\lambda_c}^{+\infty} \left( \frac{Z_D^2 e^2}{16\pi^2 \varepsilon_0^2 r^4} \right) n_D 4\pi r^2 dr \\
        & + \quad \int_{\lambda_c}^{+\infty} \left( \frac{Z_{He}^2 e^2}{16\pi^2 \varepsilon_0^2 r^4} \right) n_{He} 4\pi r^2 dr
    \end{split}
    \\
	&= \frac{n_D Z_D^2 e^2}{4\pi \varepsilon_0^2} \int_{\lambda_c}^{+\infty} \frac{1}{r^2} dr + \frac{n_{He} Z_{He}^2 e^2}{4\pi \varepsilon_0^2} \int_{\lambda_c}^{+\infty} \frac{1}{r^2} dr \\
	&= \frac{e^2}{4\pi \varepsilon_0^2 \lambda_c} \left( n_D Z_D^2 + n_{He} Z_{He}^2 \right)
\end{align}
define $f_{He}$, density proportion of $\ce{D-^3He}$:
\begin{align}
	\langle | \vec{E}_{tot} |^{2} \rangle &= \frac{e^{2}}{4\pi \varepsilon_{0}^{2}\lambda_{c}} ((n_{i}-n_{He})Z_{D}^{2}+n_{He}Z^{2}_{He}) \\
	&=\frac{e^{2}}{4\pi \varepsilon_{0}^{2}\lambda_{c}}(n_{i}Z_{D}^{2}+n_{He}(Z^{2}_{He}-Z_{D}^{2})) \\
	&=\frac{e^{2}}{4\pi \varepsilon_{0}^{2}\lambda_{c}}n_{i}(Z_{D}^{2}+f_{He}(Z_{He}^{2}-Z_{D}^{2})) \\
	&=\frac{e^{2}}{4\pi \varepsilon_{0}^{2}\lambda_{c}}  \frac{n_{i}}{2}(Z_{D}^{2}+Z_{He}^{2}) \qquad 50:50 \ fuel
\end{align}

Then:
\begin{equation}
	a^{2}=\frac{e^{4}}{4\pi \varepsilon_{0}^{2} m_{e}^{2}\lambda_{c}} \frac{n_{i}}{2}(Z_{D}^{2}+Z_{He}^{2})
\end{equation}
\begin{equation}
	P_{1e^{-}}= \frac{e^{6}}{24\pi^{2} \varepsilon_{0}^{3}m_{e}^{2}c^{3}\lambda_{c}} \frac{n_{i}}{2}(Z_{D}^{2}+Z_{He}^{2})
\end{equation}
\begin{equation}
	P_{b}= \frac{e^{6}}{24\pi^{2} \varepsilon_{0}^{3}m_{e}^{2}c^{3}\lambda_{c}} \frac{n_{e}n_{i}}{2}(Z_{D}^{2}+Z_{He}^{2})
\end{equation}

Electronic density is slightly different from D--T reaction:
\begin{equation}
n_{e}=Z_{D}n_{D}+Z_{He}n_{He}=\frac{n_{i}}{2}(Z_{D}+Z_{He})
\end{equation}

Finally:
\begin{equation}
	\boxed{P_{b}(T_{e})=\frac{n_{i}^{2}e^{6}}{96\pi^{2} \varepsilon_{0}^{3}m_{e}^{3/2}c^{3}\hbar}(Z_{D}^{2}+Z_{He}^{2})(Z_{D}+Z_{He})\sqrt{ T_{e} }}
\end{equation}

\bibliography{references}

\end{multicols}

\end{document}